\documentclass[]{aa}  
\usepackage{amsmath}
\usepackage{amssymb}
\usepackage{natbib}
\usepackage{graphicx}
\usepackage{txfonts}
\usepackage[english]{babel}
\usepackage{hyperref}
\usepackage{color}
\usepackage{units}
\begin{document}
\title{The abundance and thermal history of water ice in the disk surrounding HD\,142527 from the DIGIT Herschel Key Program}


\author{
	M. Min\inst{1,2}
		\and
	J. Bouwman\inst{3}
		\and
	C. Dominik\inst{2}
		\and
	L.~B.~F.~M. Waters\inst{1,2}
		\and
	K.~M. Pontoppidan\inst{4}
		\and
	S. Hony\inst{5}
		\and
	G.~D. Mulders\inst{6}
		\and
	Th. Henning\inst{3}
		\and
	E.~F. van Dishoeck\inst{7}
		\and
	P. Woitke\inst{8}
		\and
	Neal~J. Evans~II\inst{9}
		\and
	The~DIGIT~Team
}

\authorrunning{Min et al.}
\titlerunning{The abundance and thermal history of water ice in the disk surrounding HD\,142527}

\offprints{M. Min, \email{M.Min@uva.nl}}

\institute{
SRON Netherlands Institute for Space Research, Sorbonnelaan 2, 3584 CA Utrecht, The Netherlands
	\and
Astronomical institute Anton Pannekoek, University of Amsterdam, Science Park 904, 1098 XH, Amsterdam, The Netherlands
	\and
Max-Planck-Institute for Astronomy, K\"onigstuhl 17, 69117 Heidelberg, Germany
	\and
Space Telescope Science Institute, 3700 San Martin Drive, MD 21218, Baltimore, USA
	\and
Universit\"{a}t Heidelberg, Zentrum f\"{u}r Astronomie, Institut f\"{u}r Theoretische Astrophysik, Albert-Ueberle-Str. 2, 69120 Heidelberg, Germany
	\and
Lunar and Planetary Laboratory, The University of Arizona, Tucson, AZ 85721, USA
	\and
Leiden Observatory, PO Box 9513, NL-2300 RA Leiden, The Netherlands
	\and
SUPA, School of Physics \& Astronomy, University of St Andrews, North Haugh, St Andrews KY16 9SS, UK
	\and
Department of Astronomy, The University of Texas at Austin, 2515 Speedway, Stop C1400, Austin, TX 78712-1205, USA
}

   \date{\today}

 
  \abstract
   {The presence or absence of ice in protoplanetary disks is of great importance for the formation of planets. By enhancing the solid surface density and increasing the sticking efficiency, ice catalyzes the rapid formation of planetesimals and decreases the time scale for giant planet core accretion.}
   {In this paper we analyse the composition of the outer disk around the Herbig star HD~142527. We focus on the composition of the water ice, but also analyse the abundances of previously proposed minerals.}
   {We present new Herschel far infrared spectra and a re-reduction of archival data from the Infrared Space Observatory (ISO). We model the disk using full 3D radiative transfer to obtain the disk structure. Also, we use an optically thin analysis of the outer disk spectrum to obtain firm constraints on the composition of the dust component.}
   {The water ice in the disk around HD~142527 contains a large reservoir of crystalline water ice. We determine the local abundance of water ice in the outer disk (i.e. beyond 130\,AU). The re-reduced ISO spectrum differs significantly from that previously published, but matches the new Herschel spectrum at their common wavelength range. In particular, we do not detect any significant contribution from carbonates or hydrous silicates, in contrast to earlier claims.}
   {The amount of water ice detected in the outer disk requires $\sim80\,$\% of the oxygen atoms. This is comparable to the water ice abundance in the outer solar system, in comets and in dense interstellar clouds. The water ice is highly crystalline while the temperatures where we detect it are too low to crystallize the water on relevant time scales. We discuss the implications of this finding.}


   \maketitle
%

\section{Introduction}

Ice is an important constituent in many planet formation scenarios. Ice coated grains are expected to stick together much more easily, and the enhanced surface density of solid material in the regions where water ice can exist makes forming larger structures easier. The increase of the mass in solid material in the region where there is ice compared to where there is no ice has been estimated to range from a factor of 1.6 \citep[see e.g.][]{2011Icar..212..416M} to a factor of 4.2 \citep[see e.g.][]{2006plfo.book..129T}. Direct observations of the icy regions in planet forming disks and the abundance of ice in these regions are sparse. From an analysis of the ISO LWS spectra of a collection of protoplanetary disks, \citet{2001ApJ...547.1077C} conclude that the ice features detected from two sources \citep[confirmed by][]{2002A&A...385..546C} can be reproduced by assuming 50\,\% of the oxygen atoms are locked away in water ice.

HD 142527 is classified as a Herbig Ae/Be system. It consists of
an F-type star surrounded by a large protoplanetary disk. The disk, which extends out to beyond $1$\arcsec\ in scattered light as well as in millimeter continuumn (which translates to 145\,AU for the 145\,pc distance to the source), basically consists of a large ring with an inner radius around $\sim 140\,$AU and a heavily depleted region inside \citep{2012A&A...546A..24R, 2013A&A...556A.123C, 2014ApJ...781...87A}. Although the origin of the gap is uncertain, a likely possibility is the presence of a multi-planet system. 

When spatially resolved, the disk is seen to be highly asymmetric in both cool dust emission \citep{2013Natur.493..191C} and in rotational molecular lines \citep{2014arXiv1407.1735V}. This is attributed to azimuthal drift of large dust grains accumulated in the ring, triggered by azimuthal variations in gas mass. These so-called 'dust traps' are speculated to be important for planet formation \citep{1997Icar..128..213K, 2013A&A...550L...8B}. Multiple recent imaging studies have brought interesting features of the outer disk to light. For example, imaging polarimetry has revealed the presence of spiral arms \citep{2013A&A...556A.123C, 2014ApJ...781...87A}, and the presence of an inner disk with a different inclination than the outer disk \citep{2015ApJ...798L..44M}. The spiral arms are potentially formed by resonances with planets inside the gap, or by gravitational instabilities in the outer disk. They have also been detected in rotational CO lines, and an analysis of their characteristics suggest that they are formed by a combination of these mechanisms \citep{2014ApJ...785L..12C}. To add to the intriguing complexity of the system, \citet{2012ApJ...753L..38B} detect a companion with a separation of $~12\,$AU \citep[see also][for a confirmation of this object]{2014ApJ...781L..30C} and a mass around 0.13\,$M_{\sun}$ \citep{2015arXiv151109390L}.

The disk surrounding HD~142527 has an interesting mineralogy. The composition of the innermost few AU of the disk has a very large fraction of crystalline silicates, while further out there is a larger fraction of amorphous silicates, similar to those in the interstellar medium \citep{2004Natur.432..479V}. 

A mineralogical analysis of the ISO SWS and LWS spectrum, from 2 to 200\,$\mu$m, was presented in \citet{1999A&A...345..181M}. They detect crystalline water ice along with a broad feature around 105\,$\mu$m which they attributed to the hydrous silicate montmorillonite. Later, similar features were identified in the ISO spectra of other sources and attributed to calcite or other forms of hydrous silicates \citep{2005A&A...432..547C}. \citet{2008A&A...492..117M} disputed the identification of montmorillonite as the carrier of the broad 105\,$\mu$m band because at low temperatures the montmorillonite band is much narrower than the putative feature identified by \citet{1999A&A...345..181M}. A detailed analysis of the mineralogy with spectra from the Spitzer space telescope, from 5 to 35\,$\mu$m, is presented in \citet{2010ApJ...721..431J}.

In this paper we report on the detection of water ice with the Herschel Space Observatory in the framework of the Dust, Ice, and Gas In Time (DIGIT) Herschel Key Program \citep[][]{2016AJ....151...75G}. Crystalline water ice in protoplanetary dust was tentatively detected before with Herschel in the disk around GQ Lup \citep{2012ApJ...759L..10M}, and in a later paper also in a somewhat larger sample of T Tauri disks \citep{2015ApJ...799..162M}. Here we present an unambiguous detection of crystalline water ice in the disk around HD~142527. We determine the amount of ice needed to explain the spectroscopic features and compare to the elemental oxygen abundance. In addition, we revisit the ISO spectra to check for the previously claimed features and to combine the long wavelength analysis of the ice with Herschel with the strong 43\,$\mu$m ice feature.

The paper is organized as follows. First we introduce the new Herschel observations we secured for this source and the recalibration of the ISO data in Sect. \ref{sec:obs}. Next we outline our modeling setup in Sect.~\ref{sec:model}. The results of fitting the new data are presented in Sect.~\ref{sec:results}, followed by a discussion of the implications in Sect.~\ref{sec:discussion}.

\section{Summary of the observations}
\label{sec:obs}

\subsection{Herschel observations}

HD~142527 was observed with the \emph{Herschel} photodetector Array Camera and Spectrometer \citep[PACS;][]{Poglitsch2010} on March 16, 2011.
These observations with the unique observation identifiers (OBSID) 1342216174 and 1342216175 were taken as part of the Dust, Ice and Gas in Time (DIGIT) Herschel key programme. The PACS instrument consists of a 5$\times$5 array of 9.4\arcsec$\times$9.4\arcsec\ spatial pixels (here after referred to as spaxels) covering the spectral range from $\sim$55-200~$\mu$m with $\lambda$/$\delta\lambda$ $\sim$ 1000-3000. The spatial resolution of PACS, which is diffraction limited, ranges from $\sim$9\arcsec\ at 55~$\mu$m to $\sim$18\arcsec\ at 200~$\mu$m. All DIGIT targets are observed in the standard range-scan spectroscopy mode with a grating stepsize corresponding to Nyquist sampling \cite[see][for further details on PACS observing modes]{Poglitsch2010}.  

The PACS observations were made with four up/down scans of the grating  and one nod cycle, and used a small chopper throw. Our data were processed through the standard processing provided by the Herschel science center up to level 0. As a next step we processed our data using the Herschel Interactive Processing Environment \citep[HIPE][]{Ott2010} with track number 13.0, build 3253, using calibration version 65 and standard interactive pipeline scripts. The infrared background emission was removed using two chop-nod positions 1.5\arcmin\ from the source in opposite directions. Absolute flux calibration was made by normalising our spectra to the emission from the telescope mirror itself as measured by the off-source positions, and a detailed model of the telescope emission available in HIPE. For a well centered source, the central spaxel contains the largest fraction of the source flux and thus provides the highest signal to noise ratio (SNR) spectra. However, small pointing error drifts of the telescope can lead to flux losses and spectral artifacts. This is especially bad when analyzing (broad) dust emission bands, as such spectral artifacts can lead to spurious results. To mitigate this we used a two fold approach: Using the PACS beam profiles we determined the source position in time and corrected possible flux losses from small pointing drifts. 
We then extracted the spectra using the central 3$\times$3 spaxels to
be sure to recover the total flux and to minimize any spectral artifacts introduced by pointing drifts of the telescope.  For well pointed sources,  the highest SNR is reached in the central spaxel. However, a single spaxel spectrum also suffers more strongly from pointing effects than the combined 3x3 spectrum. To guarantee the the best absolute flux calibration with the highest SNR spectra, we scale the continuum fluxes of the central spaxel spectrum to the continuum of the  3$\times$3 spaxel spectrum using a smoothing filter with a width of a few microns.
 Spectral rebinning was done with an oversampling of a factor of two and an upsampling of a factor of one corresponding to Nyquist sampling. Absolute flux calibration and corrections for intrumental artifacts not corrected for by the model for the  emission from the telescope mirror were done using the mean spectral response function based on repeated observations  of Ceres. Using these data we estimate that the repeatability error on the PACS spectroscopic observations is on the order of 2~\%. The error on the absolute flux calibration is estimated to be about 4~\% \cite[see also][for further details on flux calibration using asteroids]{Mueller2014}.

Besides the PACS spectra, we also obtained SPIRE spectra from Herschel for the longer wavelengths. These spectra were previously presented by \citet{2014MNRAS.444.3911V}.

\subsection{ISO LWS and SWS recalibration}

We have re-reduced the available Infrared Space Observatory (ISO)
\citep{1996A&A...315L..27K} Short Wavelength Spectrometer (SWS)
\citep{1996A&A...315L..49D} and Long Wavelength Spectrometer (LWS)
\citep{1996A&A...315L..38C} spectra of HD~142527 (PI
Waelkens). The reasons for re-reducing these data are, first,
obtaining a consistent data set reduced entirely with the most recent
version of the reduction pipelines and, second, removing biases in the overall
shape of the continuum due to the way the different orders of the
spectra are being combined to obtain a continuous SED.

HD~142527 was observed using the SWS spectrograph on
29-Feb-1996 (TDT10402046) in full range grating scanning mode (AOT01,
speed 2), covering the wavelength range from 2.3 to 45\,$\mu$m at
effective spectral resolving powers ($\lambda/\Delta\lambda$) between
300$-$600. The data have been reduced in the SWS Interactive Analysis
environment \citep[OSIA,][]{1996A&A...315L..49D} using the standard
pipeline (OLP version 10) up to the Auto-Analysis Result stage, which
yields separate flux and wavelength calibrated spectra for each
individual detector and order. We have applied an extensive manual bad
data removal by clipping data clearly affected by glitches and
baseline jumps. The cleaned spectra per detector are flat fielded,
i.e. corrected for baseline differences to agree with the mean
spectrum of all detectors. We applied offset corrections in case the
median flux of the band is less than $\sim$20\,Jy, or else multiplicative
corrections were applied. Finally, the cleaned and flat fielded spectra
from both observations are combined and rebinned on a common
wavelength grid ($\lambda/\Delta\lambda$=500). Prior to combining the
rebinned spectral orders we have shifted band 2A (4.1$-$5.3$\mu$m) by
0.5\,Jy to agree with the neighbouring bands.

\begin{figure}[!t]
\centerline{\resizebox{0.95\hsize}{!}{\includegraphics{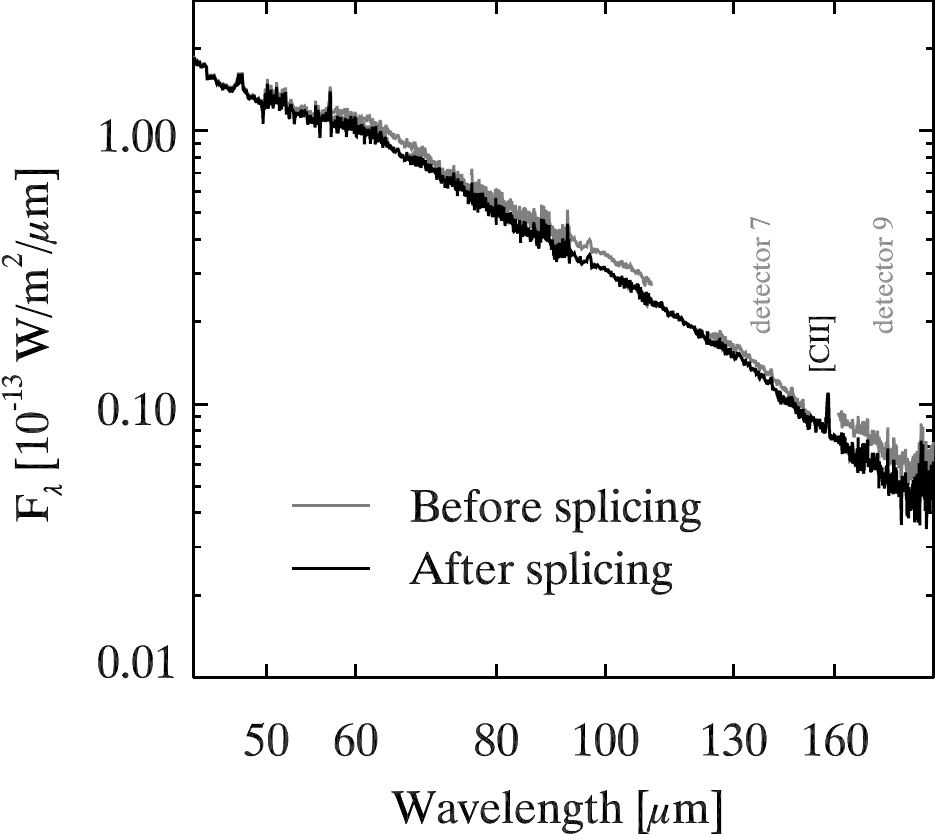}}}
\caption{HD~142527 ISO LWS spectra before and after splicing. In general the difference before and after splicing is small. We indicate the data from detectors 7 and 9 which have been shifted down by $-6$ and $-14$\,Jy, respectively, in order to form a continuous spectrum. The overall shape from 45 to 200$\,\mu$m of the dust emission is unaffected by the splicing we have applied. The only line emission which is significantly detected is the [C{\sc ii}] at 158$\,\mu$m.}
\label{fig:splicing}
\end{figure}

The LWS spectrum (45$-$200\,$\mu$m) was also obtained on 29-feb-1996
(TDT10402250) in the full range scanning mode (LWS01). We started from
the LWS Auto-Analysis Result produced by the standard LWS pipeline
(OLP version 10). Further reduction consisted of extensive bad data
removal using the ISO Spectral Analysis Package and rebinning on a
fixed resolution grid of $\lambda/\Delta\lambda$=50. The rebinned
spectra exhibit significant offsets between data from different
spectral orders, called detectors. We have corrected for
these offsets in such a way that the final spliced spectra conserved
as much as possible the overall slope (as measured in log(F$_\nu$) as
a function of log($\lambda$)) and shape of the broad energy
distribution (see Fig. \ref{fig:splicing}). No corrections have been
applied to detectors 2 and 8, detectors 0, 1 and 3 required a small
($<$4\%) correction, while detectors 2, 4 and 5 needed to be scaled
down by $\sim$10\%. The baselines of detectors 7 and 9 are clearly
discrepant. The offsets applied to these data, are -6 and -14 Jy,
respectively.

Note, that this approach for creating a continuous spectrum is
conservative in the sense that we take care not to introduce broad
features or slope changes -- which are not present in the data prior
to combining -- while ``splicing'' these independent orders. This
methodology is different from that employed in an earlier reduction
\citep{1999A&A...345..181M} where it was assumed that jumps between
data from different detectors in LWS were due to inaccurate removal
of dark currents, the correction for which can lead to significant
changes in the spectral slope. 
The SWS and LWS data did not need to be further scaled to form a
single continuous spectrum from 2.3 to 200 micron.

The final shape of the ISO spectrum is significantly different from earlier reductions. The feature previously reported around 90\,$\mu$m is not detected in the recalibrated spectrum (see also Section~\ref{sec:llf}). We identify calibration issues with the ISO LWS spectrum as the reason for a misidentification.

\begin{figure}[!t]
\centerline{\resizebox{\hsize}{!}{\includegraphics{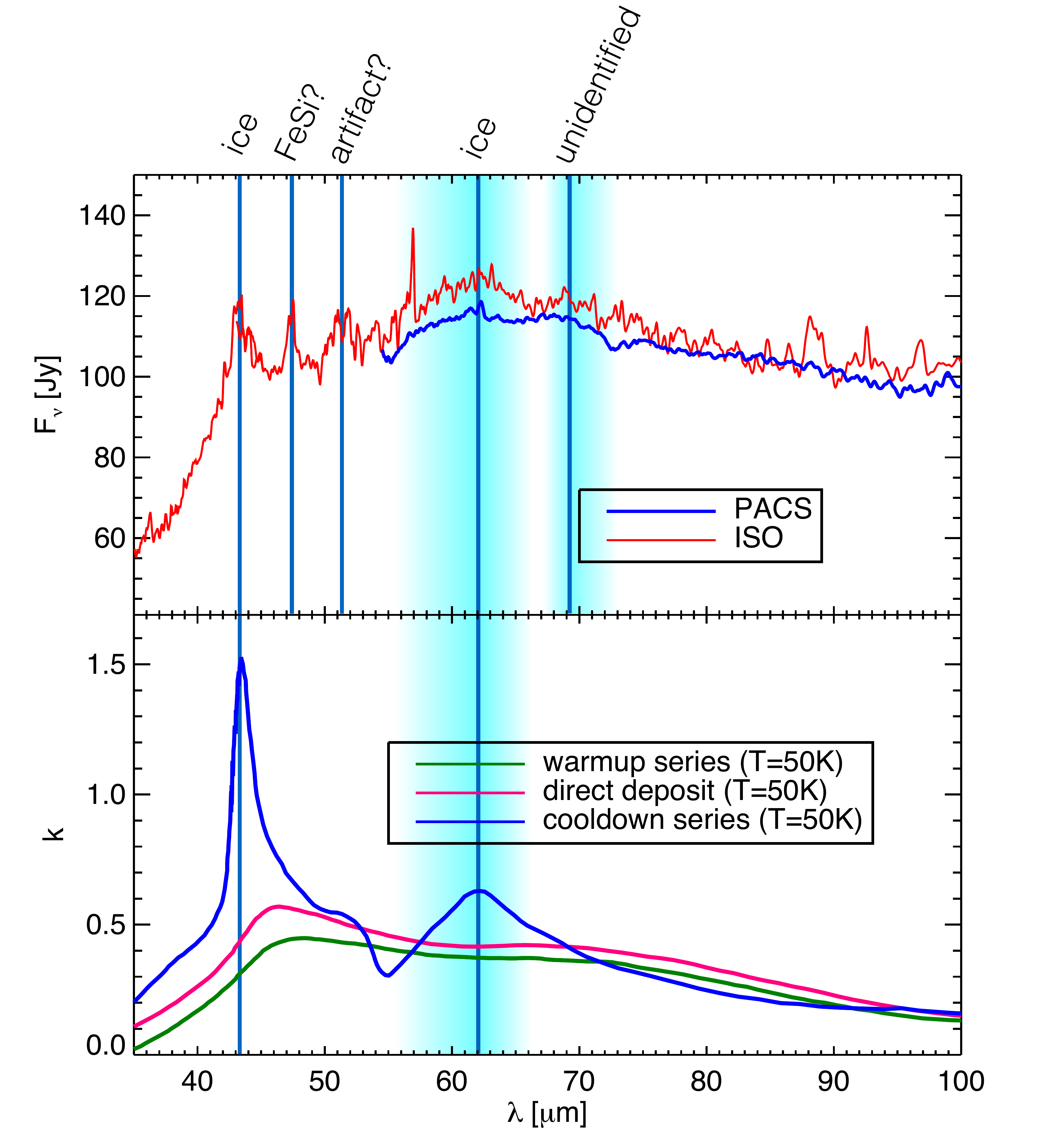}}}
\caption{The final Herschel/PACS and ISO spectra for HD~142527 (upper panel) rebinned at a spectral resolution of $R=300$ and the imaginary part of the refractive index for the different measurement series of water ice at a measurement temperature of 50\,K (lower panel). Indicated are the spectral locations of the different features visible in the spectrum. Only the cooldown series displays the crystalline ice features observed in the PACS and ISO spectra.}
\label{fig:Minerals}
\end{figure}

\subsection{Compositional inventory}

First we will analyse the long wavelength spectra directly. We will not go into the mineralogy of the inner regions, probed by the Spitzer and ISO SWS/LWS spectra. For a compositional analysis of this spectral range we refer to \citet{2005A&A...437..189V, 2010ApJ...721..431J}.

In the top panel of Fig.~\ref{fig:Minerals} we plot the resulting Herschel and ISO spectra. Very prominent are the two features of crystalline water ice at 43 and 62\,$\mu$m. In addition to these features there is a sharp feature in the ISO spectrum at 47\,$\mu$m. A suggestion for the origin of this feature from the literature is FeSi \citep{2000A&A...357L..13F}, but a firm identification, including other features (for example the expected FeSi feature around 32\,$\mu$m is not detected in the Spitzer spectrum), is currently lacking. In addition, the condensation calculations performed by \citet{2000A&A...357L..13F} that predict the presence of FeSi have been performed for evolved S-stars and it is in principle not expected for the elemental composition in a protoplanetary disk. There is structure in the ISO spectrum around 51\,$\mu$m, but it is unclear if this is simply a noisy part of the spectrum or if there is really a solid state feature. Finally, a broad feature can be seen in both the ISO and the PACS spectrum around 69\,$\mu$m. We have not been able to find an identification for this feature, but since it is visible in both the ISO and the PACS spectra, we conclude it is real. Forsterite has a feature at this wavelength, but this feature is much narrower and cannot explain the broad feature detected even when we would consider a wide distribution in iron content and temperatures. The crystalline silicate diopside, CaMgSi$_2$O$_6$, has a feature with the correct width around this wavelength position, but peaking at $65\,\mu$m \citep{2001PASJ...53..243C}. It is therefore tempting to speculate that it is caused by a mineral with composition close to diopside, but perhaps with Fe instead of Mg causing the feature to shift towards the red \citep[similar to what happens for crystalline olivine, ][]{2001A&A...378..228F}.

In Section~\ref{sec:llf} we will perform a more detailed compositional analysis including the previously proposed materials on the basis of the ISO spectrum.

\section{Modeling setup}
\label{sec:model}

Our main aim in this study is to constrain the composition and the
location of the water-ice that gives rise to the observed bands in the
Herschel/PACS spectrum. These ice bands arise from the cooler, outer
parts of the disk. However, in order to derive reliable parameters of
the ice one needs a full, realistic model of the structure of the disk
including the ice free inner parts. 
Our model partly builds on previous detailed modeling of the HD 142527 system \citep{2011A&A...528A..91V}. Recently it was identified that the inner and the outer disk of HD~142527 are not aligned \citep{2015ApJ...798L..44M}. This causes the shadow of the inner disk on the outer disk wall to be significantly smaller than in previous models, where the inner disk is aligned. This also reflects itself in the temperature distribution of the inner edge of the outer disk. Therefore, we cannot ignore this effect in our modelling of the outer disk and we adopt a full 3D model including an inclined inner disk.

We adopt a two step modeling approach. First, we improve on the full radiative transfer model setup by \citet{2011A&A...528A..91V} by using a 3D model setup in combination with an automatic genetic fitting algorithm to optimize the model parameters, taking into account the updated spectral energy distribution (SED) and the new Herschel data. In the second step we take the resulting size distribution of the dust grains to make a least squares fit to a model of optically thin emission from the spectral region with the ice features. In this second step we assume that the thermal emission features are created in an optically thin warm surface layer and make a linear least square fit to a part of their spectrum assuming a single emission temperature for the feature contributing materials and a continuum made of blackbody emission. The advantage of this two step approach is that we have a relatively robust estimate of the size distribution of the grains from the multi-wavelength radiative transfer model and a detailed analysis of the mineralogy from the optically thin least squares analysis (see Section~\ref{sec:llf} for details).

\subsection{Multi-wavelength radiative transfer modeling}

As our starting point we take the model from \citet{2011A&A...528A..91V} with significant modifications. The basis of this model is a passive irradiated disk. We include an inner disk, from the dust evaporation radius up to 13 AU, and an outer disk, starting around 130\,AU. The vertical structure of the gasdisk is parametrized using a scale height which varies as a powerlaw of the distance to the star. We adopt a tunable, size dependent settling prescription based on self-consistent considerations. Each grain size is settled to its own scale height according to the settling prescription as described in \citet{2015arXiv151103431W}, which is basically the prescription by \cite{1995Icar..114..237D} with a simple adjustment for the parameterized vertical gas structure. In the description the sedimentation is related to a single free parameter, the turbulent $\alpha$.

We construct a full radiative transfer model of the disk to compute the temperature structure and determine whether water ice is present at a given location.  For this we use the temperature-dependent opacity option in the 3D version of the continuum radiative transfer program MCMax \citep{2009A&A...497..155M}. This allows us to include the temperature dependent ice optical properties as measured by \citet{1994MNRAS.271..481S} (see also next section). The temperature dependent opacity implementation was successfully used in e.g. \citet{2011A&A...531A..93M}. 

\begin{table}[!tb]
\caption{The composition of the dust mixture and references to the laboratory data used.}
\begin{center}
\begin{tabular}{lc|l}
\multicolumn{2}{l|}{name}		&	reference	\\
\hline
\multicolumn{3}{c}{silicate component \& mass fractions} \\
\hline
amorphous pyroxene		&	73.2\,\%		&	{\cite{1995A&A...300..503D}}	\\
crystalline forsterite			&	8.6\,\%		&	{\cite{1973PSSBR..55..677S}}	\\
crystalline enstatite			&	14.8\,\%		&	{\cite{1998A&A...339..904J}}	\\
amorphous silica			&	3.4\,\%		&	{\cite{1960PhRv..121.1324S}}	\\
\hline
\multicolumn{3}{c}{other components}\\
\hline
\multicolumn{2}{l|}{amorphous carbon}	&	{\cite{1993A&A...279..577P}}	\\
\multicolumn{2}{l|}{water ice}			&	{\cite{1994MNRAS.271..481S}}	\\
\multicolumn{2}{l|}{montmorillonite}		&	{\cite{1990MNRAS.246..332K}}	\\
\multicolumn{2}{l|}{serpentine}			&	{\cite{1990MNRAS.246..332K}}	\\
\multicolumn{2}{l|}{chlorite}			&	{\cite{1990MNRAS.246..332K}}	\\
\multicolumn{2}{l|}{dolomite}			&	{\cite{2007ApJ...668..993P}}	\\
\multicolumn{2}{l|}{calcite}				&	{\cite{2007ApJ...668..993P}}	\\
\end{tabular}
\end{center}
\label{tab:opticaldata}
\end{table}

The radial density distribution of the dust disk was parameterized using a radial surface density \citep{2008ApJ...678.1119H}
\begin{equation}
\Sigma(r) \propto R^{-q}\exp \left\{- \left( \frac{R}{R_\mathrm{exp}} \right) ^{2-q}\right\},
\end{equation}
for $R_\mathrm{in} < R < R_\mathrm{out}$. Here $R_\mathrm{exp}$ is the turnover
point beyond which an exponential decay of the surface density sets in
and $q$ sets the power law in the inner region. Although we fix the outer radius of the disk at 2000\,AU, the surface density of the disk is already highly diminished after the turnover radius which is a fitting parameter with a typical value around a (few) hundred AU.

We use a two-component model consisting of an inner disk and an outer disk. The inner disk defines the SED up to a wavelength of $\sim15\,\mu$m. This part of the SED was very difficult to understand using only an inner disk in hydrostatic equilibrium, especially because of the large near-infrared excess \citep[see][for extensive discussion on these components to the modeling]{2011A&A...528A..91V}, therefore a halo component was invoked which may represent a vertically extended disk wind \citep[e.g.][]{2012ApJ...758..100B}. Here, we have used a parameterized structure of the inner disk and thus find a solution without a halo, but an inner disk which is significantly extended beyond hydrostatic equilibrium. Obtaining the exact structure of the inner regions is not the aim of this paper, but it could be constrained by infrared interferometric observations. Here we use the inner disk model to get a decent fit to the SED and shadowing effects on the outer disk, which we use to constrain the composition of the outer disk component.

\subsection{Dust optical properties}

We use Bruggeman effective medium theory to mix the different materials present in the dust (including the ice). In contrast, \citet{2011A&A...528A..91V} simply add the opacities of pure grains. Using effective medium theory provides a better account of the effects of mixed grain compositions and of ice on the temperature structure of the disk. We take the same material composition as \citet{2011A&A...528A..91V}, based on the mineralogical analysis of 10$\,\mu$m spectra by \citet{2005A&A...437..189V}, but add in 25\% vacuum to simulate porous grains. See Table~\ref{tab:opticaldata} for references to the refractive index data used. Like \citet{2011A&A...528A..91V}, we use the Distribution of Hollow Spheres \citep[DHS][]{2005A&A...432..909M} to compute the resulting optical properties of the grains from the refractive indices obtained from the effective medium theory. We take the irregularity parameter $f_\mathrm{max}=0.8$ to properly account for particle irregularities. We adopt a continuous size distribution $n(a)\propto a^{-p}$, with $a$ the size of the dust grains. The sizes range from $a=0.05$ to $a=3000\,\mu$m.

\subsection{The location of water ice in the disk}

To determine the location of the ice we set a limit on the temperature, density, and the strength of the local UV radiation field. For the temperature and pressure dependent part of the ice condensation and sublimation we follow the equations from \citet{2009A&A...506.1199K, 2011Icar..212..416M}. The vapor pressure, i.e. the equilibrium partial pressure at the sublimation temperature, is given by \citep[after Eq.~1 in][]{2009A&A...506.1199K},
\begin{equation}
\label{eq:Pv}
p_v(T)=\frac{k_B}{\mu m_u}\cdot10^{\left(B-A/T\right)}\cdot(1\mathrm{K})
\end{equation}
with $T$ the temperature and $A$ and $B$ the constants known from laboratory measurements \citep[$A=2827.7$\,K, $B=7.7205$;][]{1994ApJ...421..615P, 2009A&A...506.1199K}. If we assume that recondensation is proportional to the number of collisions on the surface of the grain with proportionality constant $\alpha_\mathrm{stick}$, the mass increase per unit surface area per unit time of a grain becomes,
\begin{equation}
J^+(T)=\alpha_\mathrm{stick} p \sqrt{\frac{\mu m_u}{2\pi k_B T}},
\end{equation}
where $p$ is the partial pressure of the condensing material, and $\mu m_u$ is the mass of a single molecule. At equilibrium ($p=p_v$), i.e. when the mass loss due to thermal evaporation equals the mass increase due to recondensation, the mass loss rate per unit surface area is given by $J^-(T)=J^+(T)$.

Eq.~\ref{eq:Pv} was derived without UV destruction of the ice. Assuming the ice mass loss rate due to UV photodesorption is proportional to the strength of the UV field, we get for the mass loss rate per unit surface area,
\begin{equation}
J^-(T)=\alpha_\mathrm{stick} p_v(T)\sqrt{\frac{\mu m_u}{2\pi k_B T}} + \gamma' G_{UV}
\end{equation}
where $G_\mathrm{UV}$ is the local UV field strength in terms of Habing fields \citep[$u_\mathrm{UV}^\mathrm{Hab}=5.33\cdot10^{-14}\,$erg\,cm$^{-3}$][]{1968BAN....19..421H}, and $\gamma'$ is a scaling parameter.
Now if we set $J^+(T)=J^-(T)$, we get the required minimum water vapor density for the ice to exist becomes,
\begin{equation}
\label{eq:finalTsub}
\rho(\mathrm{H_2O})=10^{\left(B-A/T\right)}\frac{1\mathrm{K}}{T}\mathrm{g/cm}^3+\gamma\,G_{UV}\sqrt{\frac{1\mathrm{K}}{T}}.
\end{equation}
\begin{equation}
\gamma=\frac{\gamma'}{\alpha_\mathrm{stick}}\sqrt{\frac{2\pi\mu m_u}{k_B}}.
\end{equation}

We use the photochemical disk modelling code ProDiMo \citep{2009A&A...501..383W} to estimate empirically at which UV field strength the ice is destroyed and determine the value of $\gamma$ to use. We find a value of $\gamma=10^{-20}$g\,cm$^{-3}$ corresponds well to the detailed chemistry and photodesorption rates computed in ProDiMo. So, for example, this means that for a typical field strength in the outer disk of $G_{UV}=100$ and a temperature of $100$\,K we need a minimum water vapor density of $\rho(\mathrm{H_2O})\approx10^{-19}$g\,cm$^{-3}$ (or $n(\mathrm{H_2O})\approx3300$\,cm$^{-3}$) for water ice to be stable, which is significantly higher than the densities typically found in the surface layer of the disk, especially in the outer regions. Using Eq.~\ref{eq:finalTsub}, we can compute the sublimation temperature of the ice locally in the disk, also taking into account the desorption caused by UV photons. We iterate the radiative transfer and ice sublimation to arrive at a consistent solution. Note that this treatment does not capture all the details of water chemistry and ice formation and destruction present in detailed chemical computations.

\subsection{Optical properties of water ice}
\label{sec:ice}

The solid state features of water ice are influenced by its formation temperature and thermal history, and by its current temperature. The formation temperature and thermal history mainly determine whether the ice is amorphous or crystalline, and thus which features are present or absent. The current temperature, i.e. the temperature at which the ice emits, determines the exact wavelength positions and shapes of these features. If the ice originally condenses in amorphous form, but is subsequently heated to temperatures above $\sim110\,$K, the ice becomes crystalline, changing its spectroscopic appearance. It stays crystalline even if subsequently cooled down.

\cite{1994MNRAS.271..481S} measured the temperature-dependent optical properties of water ice formed at different temperatures. They measured three different sets of optical constants; a set where the ice was condensed at 10\,K and warmed up to the measurement temperature (series I), a set where the condensation proceeded at the same temperature at which the optical properties were measured (series II), and a set where the ice was condensed at 150\,K and cooled down to the measurement temperature (series III). In the series I and II the ice is amorphous for temperatures lower than the crystallization temperature of $\sim110\,$K. Only in the series III measurements, where the ice was condensed in crystalline form and cooled down, is the ice crystalline at these lower temperatures. In addition, \citet{1994MNRAS.271..481S} found a shift in the position of the 43\,$\mu$m crystalline ice feature with temperature, where the feature shifts to shorter wavelengths with decreasing temperature. The refractive indices as measured by \citet{1994MNRAS.271..481S} do not directly give the right peak position for the solid state features. The peak is also dependent on the shape, size and structure of the grains under consideration, and we model this using the grain model described in the previous section. In our radiative transfer modeling we take the temperature dependence of the ice opacity into account self-consistently. The curves for the imaginary part of the refractive index of all three series at a measurement temperature of 50\,K are shown in the lower panel of Fig.~\ref{fig:Minerals}.

In our modeling procedure we pick one of the three series and use at each location in the disk the optical properties within that series that correspond to the local disk temperature. Thus, we always have three different models, one where the ice was formed at lower temperatures than currently observed, one where it was formed locally, and one where it was formed at higher temperatures.

\subsection{Fitting procedure}

We use a genetic algorithm to fit the spectral energy distribution and the interferometric observations \citep[also presented in][]{2011A&A...528A..91V}. The algorithm we designed for this is based on the \textsc{pikaia} algorithm \citep[see][]{1995ApJS..101..309C}, but revised to efficiently fit a combination of photometric and spectroscopic observations over a wide wavelength range. 
The possibility was added to the fitting procedure to evaluate all models generated during the fitting process to determine accurate error estimates for all fitting parameters. The error estimation process uses bootstrapping. We generate 1000 artificial observational datasets consisting of random deviations from the original dataset using the error bars on the data. For all these datasets we compute the $\chi^2$ that this dataset would have with each model. 
This method yields a distribution of $\chi^2$ values, commonly referred to as the F-distribution. We determine the standard deviation, $\sigma$, of the F-distribution. Next, we consider all models computed in the genetic fitting procedure that have a $\chi^2$ within $1\sigma$ of the best fit model. The spread on the model parameters that we find in all these models gives us the errors on the model parameters. Finally, at the end of the genetic run, we explore parameter space locally with a few hundred models centered around the best-fit parameters. This way we get a good estimate of the model uncertainties without having to compute a full scan of parameter space. For the bootstrapping method to work, the standard deviations on the observations are crucial. If these are too small, the variations in the 1000 artificial datasets are too small resulting in unrealistically small errors on the model parameters. We therefore multiply the errors on the observations by $\sqrt{\chi^2}$ before computing the 1000 artificial datasets. This way, when the fit is not good (i.e. $\chi^2>1$), the errors on the derived model values automatically increase. Note that the location of the best fit model in the parameter space is unaffected by this rescaling of the error bars; we still find exactly the same best fit model. The only thing that is affected is the estimate of the uncertainties of the parameters.

Defining $\chi^2$ for an observational dataset containing both high resolution spectra (ISO and Spitzer) as well as photometry is non-trivial. To compute a total $\chi^2$ we weigh the different observational datasets differently to prevent the fitting procedure from becoming too sensitive to small parts of the spectrum simply because there are many observations in a particular wavelength region. Determining the weights to get a 'good' fit to the spectrum is to a certain degree subjective, and sensitive to the problem at hand. We increase the weight of the region around the ice features, because in this study we are most interested in this part of the spectrum.

\subsection{Optically thin, mineralogical/chemical analysis}
\label{sec:llf}

From the radiative transfer modelling we get an estimate of the abundance of ice and of the grain size distribution. We use this as input in a linear least square fitting procedure. The aim of this exercise is to put firm constraints on the presence or absence of certain mineralogical or ice components in the dust. The disk is not  optically thin in the far infrared, but we assume here that the emission features originate in the optically thin surface layer of the disk and simulate the optically thick contribution with a distribution of black bodies. 

The procedure is based on that developed by \citet{2007A&A...462..667M} with a few small adjustments. The procedure is as follows:
\begin{enumerate}
\item\label{it:makemix} For each component we construct a template spectrum of thermal emission at 80\,K. The template spectrum includes the best fit mixture (silicates, amorphous carbon and water ice) and a small abundance of the component we want to test. We pick 80\,K as the representative temperature of the inner wall of the outer disk. There is some variation in the temperature of this inner wall, but as long as we pick a single temperature for all dust components, the exact value should not significantly influence the abundance estimates.
\item\label{it:makefit} We make a linear least square fit of the observed spectrum with this mixture plus a contribution from pure blackbody emission from a sum of blackbodies at different temperatures. For the blackbody-components we vary the temperatures between 10 and 1500\,K, the weights are determined from linear least square fitting.
\item\label{it:upper} To test the significance of each dust component, the abundance in step \ref{it:makemix} is increased until the fit in step \ref{it:makefit} gets worse by 1$\sigma$ according to a standard F-test. This abundance is the upper limit for this component. This is done for the silicate and the ice components in the mixture.
\end{enumerate}
Following the analysis of the previous reduction of the ISO data \citep{1999A&A...345..181M}, we test for the presence of two different types of carbonates (calcite and dolomite), and three different types of hydrous silicates (montmorillonite, serpentine and chlorite). As mentioned before, these materials have been proposed as components of the disk from analysis of a previous reduction of the ISO spectrum. Further, we test for the presence of the three different types of water ice corresponding to the different series of measurements by \cite{1994MNRAS.271..481S}. We determine the abundance of each of the different ice measurements, corresponding to different thermal histories (see Section \ref{sec:ice}), to track the thermal history of the ice.

\begin{figure}[!t]
\centerline{\resizebox{\hsize}{!}{\includegraphics{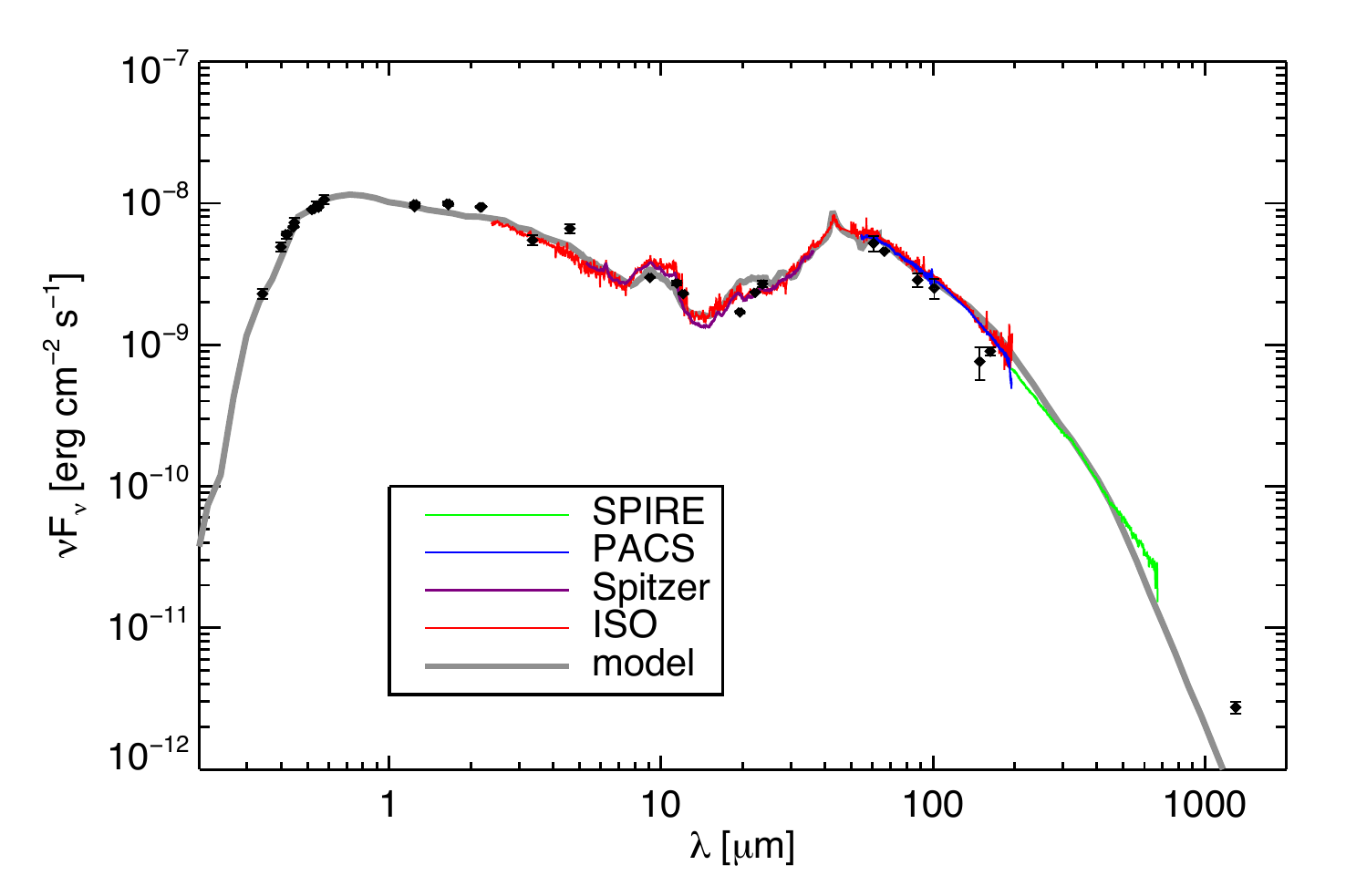}}}
\caption{The spectral energy distribution of HD~142527 together with the best fit model.}
\label{fig:SED}
\end{figure}

\begin{table}[!tb]
\caption{Parameters of the best fit radiative transfer model.}
\begin{center}
\begin{tabular}{l|c|c}
parameter					&	value		&	error \\
\hline
\multicolumn{3}{c}{Inner disk} \\
\hline
inner radius: $R_\mathrm{in}$ [AU]            	& 0.38 		& 0.09 \\
outer radius: $R_\mathrm{out}$	 [AU]			& 13			& fixed \\
dust mass: $M_\mathrm{dust}$ [M$_{\sun}$]	& $4.5\cdot 10^{-7}$  		& factor 1.6 \\
size distribution powerlaw: $p$           		& 4.9  			& 0.5 \\
surface density powerlaw: $q$        			& 1.6  			& 0.6 \\
scale height at the inner radius: $h$ [AU]		& 0.022			& 0.002 \\
scale height powerlaw: $p_h$				& 1.23			& 0.08 \\
mixing strength: $\alpha$            			& $1.3\cdot 10^{-5}$  	& factor of 2 \\
inclination: $i$							& -46$^\circ$		& 11 \\
\hline
\multicolumn{3}{c}{Outer disk} \\
\hline
inner radius: $R_\mathrm{in}$ [AU]			& 146 		& +2 / -15 \\
exponential radius: $R_\mathrm{exp}$ [AU]	& 282		& 70 \\
dust mass: $M_\mathrm{dust}$ [M$_{\sun}$]	& $1.4\cdot 10^{-3}$  		& factor 1.6 \\
size distribution powerlaw: $p$           		& 3.7  			& 0.2 \\
surface density powerlaw: $q$        			& 1.7  			& 0.5 \\
scale height at the inner radius: $h$ [AU]		& 17.3			& 2 \\
scale height powerlaw: $p_h$				& 1.1				& 0.1 \\
mixing strength: $\alpha$            			& $4.7\cdot 10^{-6}$  	& factor of 2 \\
inclination: $i$							& 20$^\circ$		& fixed \\
\hline
\multicolumn{3}{c}{Global parameters} \\
\hline
Carbon/silicate mass ratio 				& 0.2  				& 0.1 \\
Ice/silicate mass ratio 					& 1.6					& +0.9 / -0.6 \\
\end{tabular}
\end{center}
\label{tab:parameters}
\end{table}

\section{Results}
\label{sec:results}

\subsection{Properties of the disk}

The best-fit parameters of the HD 142527 disk, as derived using the genetic fitting algorithm, are summarized in Table~\ref{tab:parameters}. First we highlight one of the main findings, the ice/silicate ratio of 1.6, which is in general agreement with previously proposed values. We dedicate section \ref{sec:ice abundance} to discuss this value in detail.

From Table~\ref{tab:parameters} a few remarkable properties of the best-fit model are apparent.

First, the scale height of the inner and outer disks. At the inner edge of the inner disk, the scale height expected from hydrostatic equilibrium and a midplane temperature of 1600\,K is only 0.013\,AU, a factor of 1.7 below the bestfit value. This reflects the problem already indicated before with the near-IR excess. At the inner edge of the outer disk, the scale height expected from hydrostatic equilibrium, assuming a temperature of 80\,K is 22\,AU, much closer to the best fit value. This indicates that the outer disk is likely close to hydrostatic equilibrium.

Another interesting parameter is the turbulent mixing strength, $\alpha$, which we find to be much lower than the canonical value of $\alpha\sim10^{-2}$ to $10^{-3}$ \citep{1998ApJ...495..385H}. This is surprising since the SED of this source is very red, i.e. the disk emission peaks at very long wavelengths, which classically indicates a 'flaring' disk structure. Our model thus supports the current view that the red SED is an indication for a gapped disk structure \citep{2013A&A...555A..64M}. Finally, the total dust mass in the outer disk is relatively high. We currently assume a canonical gas-to-dust ratio of 100, which implies a total disk mass of 0.14\,M$_{\sun}$. A possibility is that the outer disk has already lost a large fraction of its gas mass. The values for the gas-to-dust ratio and the turbulence parameter, $\alpha$, are degenerate in our model. To get to the same degree of settling, but with a lower gas-to-dust ratio, we would need a higher value of $\alpha$ to keep the same coupling between gas and dust. Thus, the small value of $\alpha$ we derive under the assumption of a gas-to-dust ratio of $100$ may already indicate that the gas-to-dust ratio in reality is much lower.

In our model we have $\sim10^{-3}\,$M$_{\sun}$ of water ice in total. However, at a wavelength of $\sim50\,\mu$m the disk is not  optically thin. This implies that we cannot see the entire outer disk. To estimate the amount of mass needed to account for the observed ice features we compute the total dust mass above the $\tau_{50\mu m}=1$ surface. In our model this is approximately $3\cdot10^{-5}\,$M$_{\sun}$. If we also take into account that at this wavelength only grains smaller than $20\,\mu$m in size (i.e. $2\pi a/\lambda<1$) and warmer than $30\,$K contribute significantly to the formation of the ice features, we get a contributing mass at this wavelength of $3\cdot10^{-6}\,$M$_{\sun}$. Roughly half of this mass must be in crystalline water ice, which is about 0.5 Earth masses of ice. This is the hard lower limit on the mass in crystalline water ice needed to account for the observed infrared features. From our model we also find that, due to settling, the grain size distribution above the $\tau_{50\mu m}=1$ surface is heavily depleted in grains with sizes $a>5\,\mu$m.

\subsection{Composition of the ice and dust}

\begin{table}[!tb]
\caption[caption]{The formal upper limits of the components not significantly detected in the fitting procedure\footnotemark.}
\begin{center}
\begin{tabular}{l|c}
name					&	upper limit \\
\hline
water ice 'direct deposit'		&	40\,\% of M$_\mathrm{ice}$\\
water ice 'warmup'			&	40\,\% of M$_\mathrm{ice}$\\
\hline
montmorillonite				&	47\,\% of M$_\mathrm{dust}$\\
serpentine					&	29\,\% of M$_\mathrm{dust}$\\
chlorite					&	8\,\% of M$_\mathrm{dust}$\\
dolomite					&	3\,\% of M$_\mathrm{dust}$\\
calcite					&	2\,\% of M$_\mathrm{dust}$\\
\end{tabular}
\end{center}
\label{tab:mineralogy}
\end{table}
\footnotetext{The upper limits are derived for inclusion of single components to the best fit. For example, either the 'direct deposit' or the 'warmup' ice components can be added with an abundance of 40\,\% compared to the 'cooldown' ice component. The dust mass $M_\mathrm{dust}$ includes the mass in the crystalline ice component.}

\begin{figure}[!t]
\centerline{\resizebox{\hsize}{!}{\includegraphics{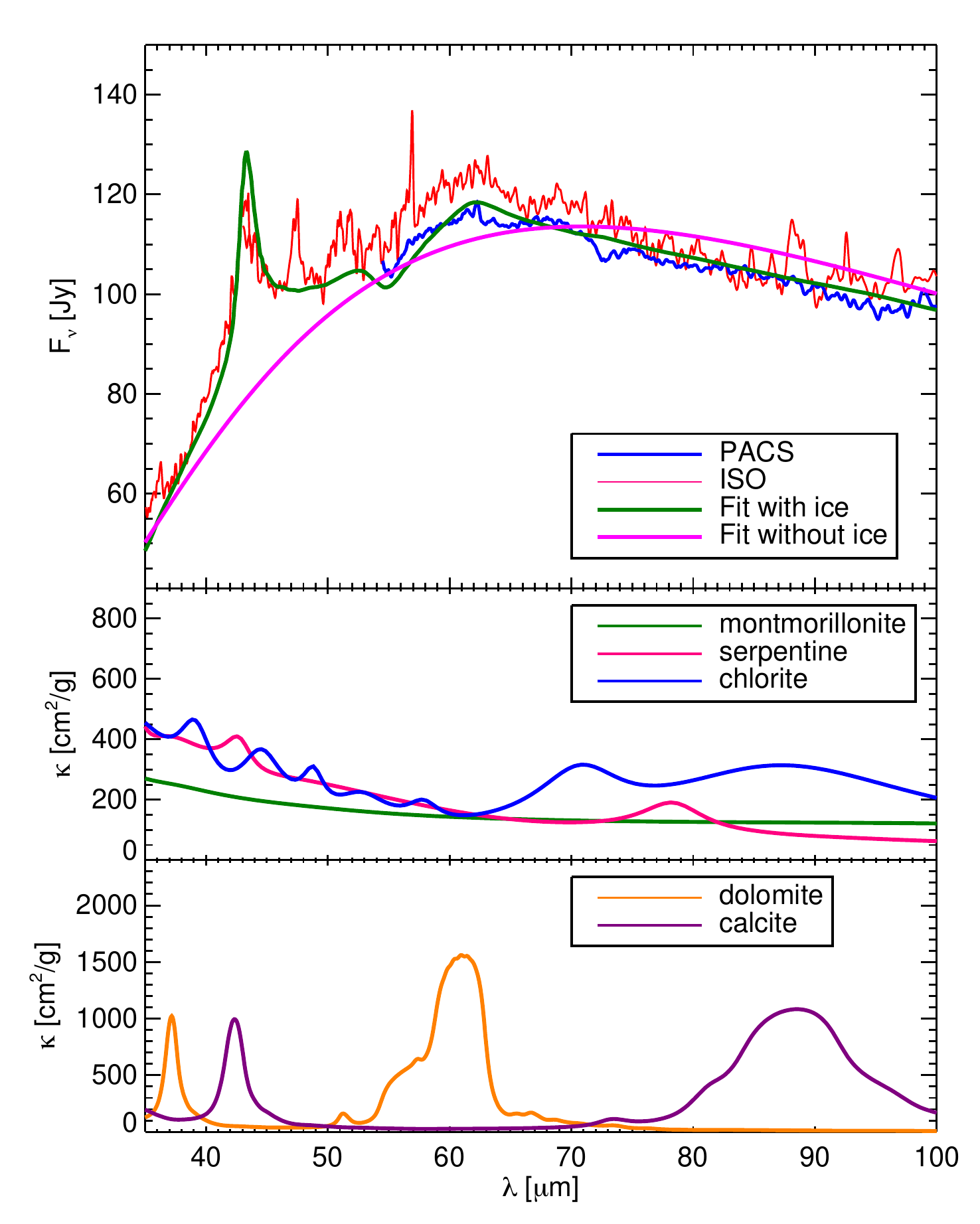}}}
\caption{Zoom in of the 35 to 100$\,\mu$m part of the spectrum together with the mass absorption coefficient of a selection of proposed materials (carbonates and hydrous silicates). In the upper panel we also display fits with and without crystalline water ice to highlight the significance of this ice component.}
\label{fig:Components}
\end{figure}

From the ice measurements of \cite{1994MNRAS.271..481S} together with our dust opacity computations, we can derive the peak position of the $43\,\mu$m feature as a function of temperature. A very simple linear relation between the measured peak position and measurement temperature is found,
\begin{equation}
\label{eq:peak43}
\lambda_\mathrm{peak}^{43\mu\mathrm{m}}=\left(42.5+\frac{T}{65\,K}\right)\mu m.
\end{equation}
Note that this relation is dependent on the dust grain model employed to convert the refractive indices into emissivities.
For the $62\,\mu$m feature we find a similar relation, though because of the large width of the feature this is not as accurate,
\begin{equation}
\label{eq:peak62}
\lambda_\mathrm{peak}^{62\mu\mathrm{m}}=\left(60.7+\frac{T}{53\,K}\right)\mu m.
\end{equation}
From the ISO spectrum we find a peak position of the ice around HD~142527 of $\lambda_\mathrm{peak}^{43\mu\mathrm{m}}=43.2\pm0.2\,\mu$m. Using Eq.~\ref{eq:peak43} this corresponds to $T=45.5\pm13\,$K. This is colder than the crystallization temperature of ice, and also colder than the temperature of the material in the gap wall (which is around 70 - 80\,K). From the PACS spectrum we find $\lambda_\mathrm{peak}^{62\mu\mathrm{m}}=62\pm0.7\,\mu$m, which corresponds to $T=69\pm37\,$K using Eq.~\ref{eq:peak62}. Because of the broadness of this feature, its spectral location is much more difficult to determine, but it is within error consistent with the low temperature derived from the 43\,$\mu$m feature.

In previous reductions of the ISO spectrum of HD~142527 one could identify a broad feature around 100\,$\mu$m. The proposed carriers of this feature were carbonates or hydrous silicates \citep{1999A&A...345..181M, 2005A&A...432..547C}. However, the feature is not detected in the PACS spectrum, consistent with a careful re-analysis of the ISO LWS spectrum. Therefore, we conclude there is no evidence for these species in this source. 

To put constraints on the abundance of carbonates and hydrous silicates in the outer disk of HD~142527 we follow the linear least square fitting procedure described in section \ref{sec:llf}. Indeed, none of the carbonates or hydrous silicates are significantly detected. Adding these components to the fitting procedure actually increases the reduced $\chi^2$. See Fig.~\ref{fig:Components} for the best fit we find. Note that this fit follows the PACS spectrum more closely than the ISO spectrum because of the superior quality of the spectrum (and thus lower error bars). For water ice, we  significantly detect only the 'cooldown' series in the spectrum. Thus, the ice in the outer disk is compatible with being fully crystalline. In Table~\ref{tab:mineralogy} we summarise the upper limits found for the other components. The high values for montmorillonite and serpentine are caused by their very weak spectral signature, which makes them act as a continuum component in this wavelength range. This means that although there is no indication for the presence of these materials in the spectrum, we cannot rule out their presence. For montmorillonite it was shown that at low temperatures (around 10\,K) the features are significantly stronger and sharper, which would reduce the upper limit on the montmorillonite \citep{2008A&A...492..117M}.

\subsection{Scattered light images}

\begin{figure}[!t]
\centerline{\resizebox{\hsize}{!}{\includegraphics{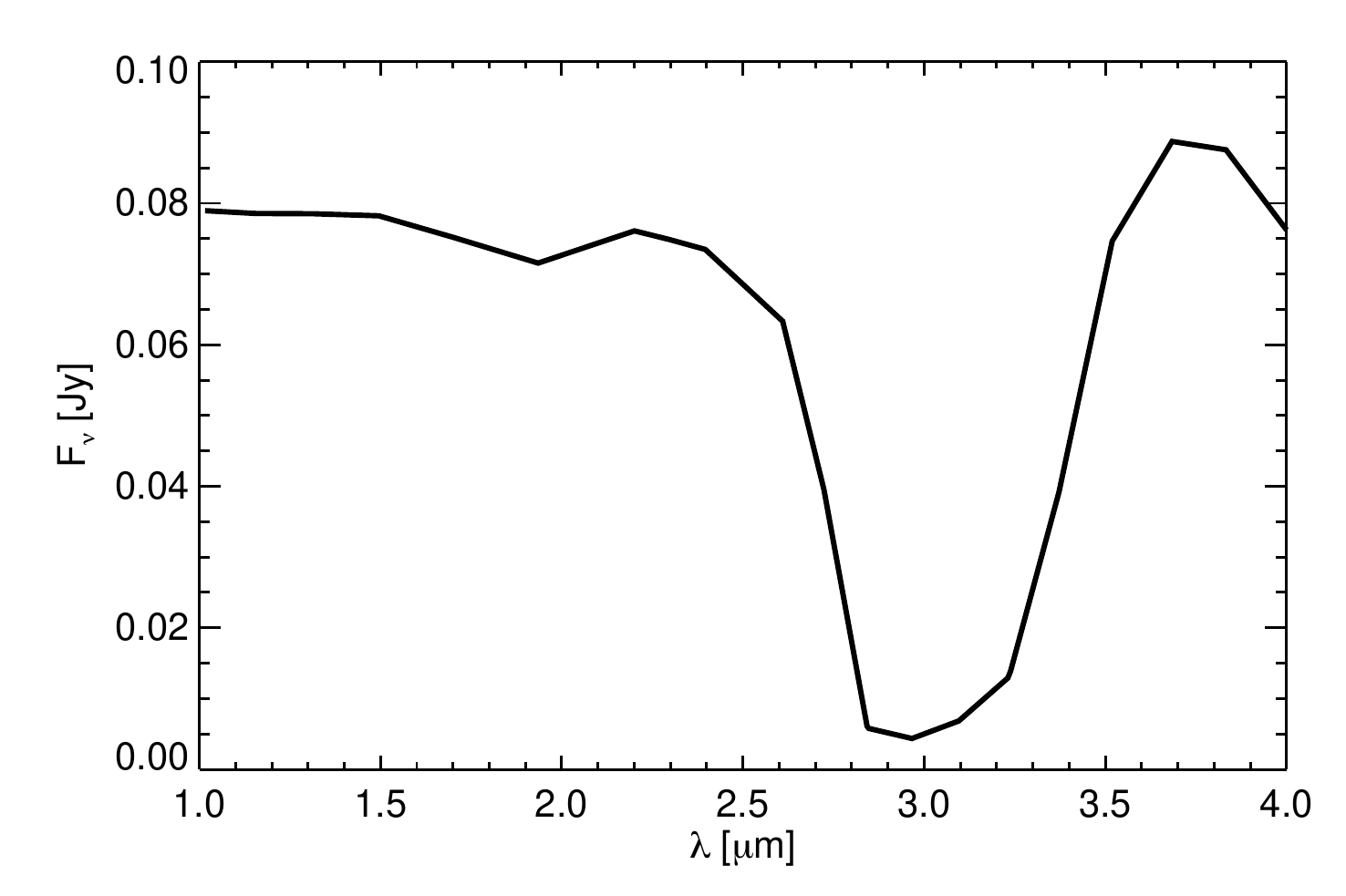}}}
\caption{The model spectrum of the outer disk. The $3\,\mu$m ice feature, as detected by \citet{2009ApJ...690L.110H} is clearly seen.}
\label{fig:OuterDisk}
\end{figure}

\begin{figure}[!t]
\centerline{\resizebox{0.9\hsize}{!}{\includegraphics{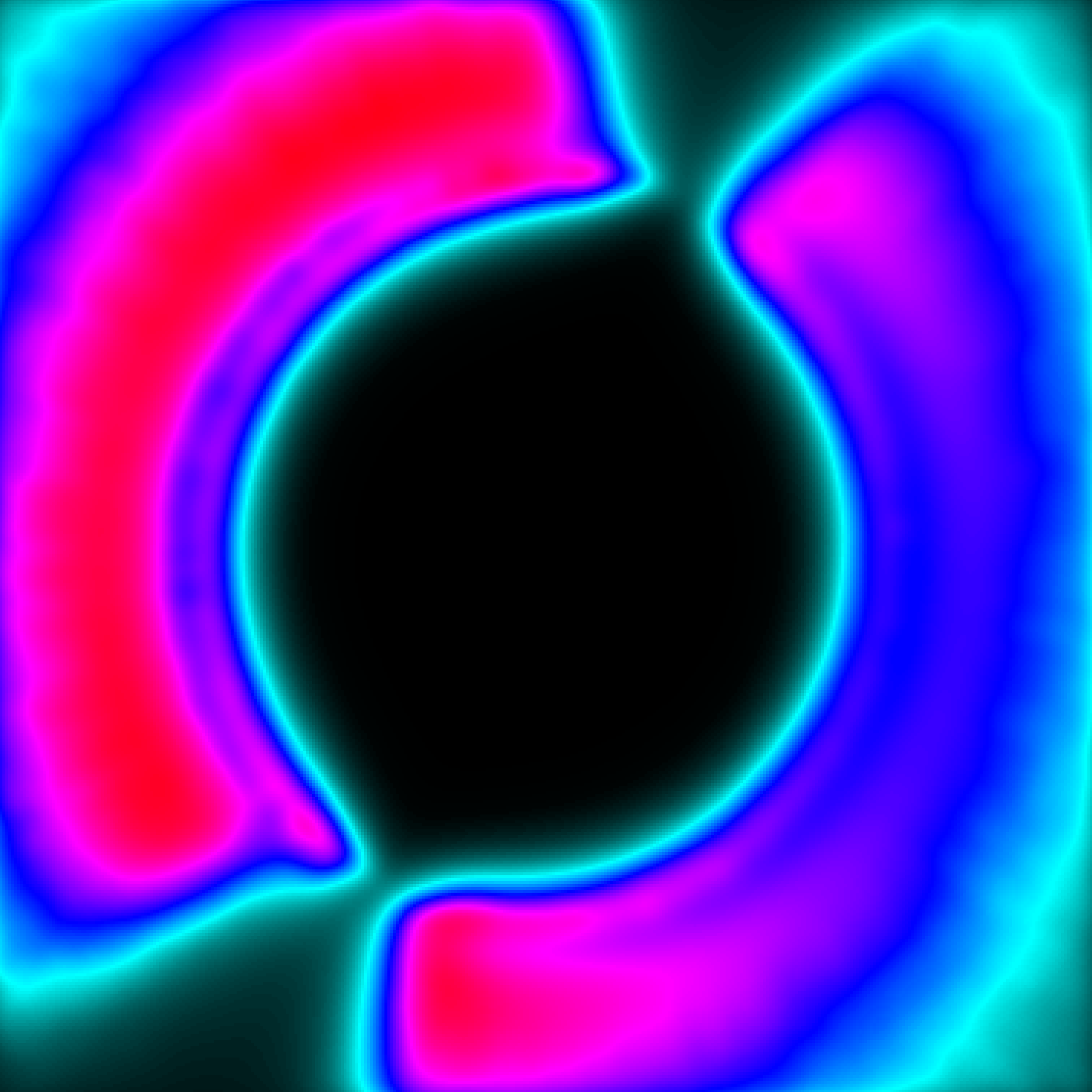}}}
\centerline{\resizebox{0.8\hsize}{!}{\includegraphics{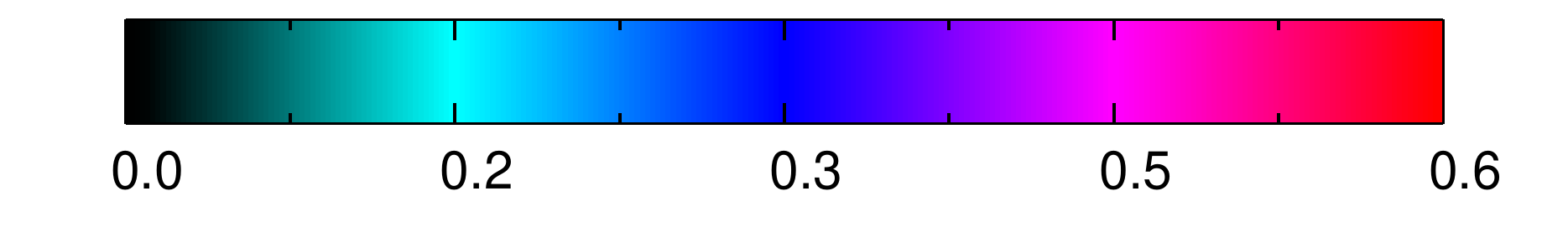}}}
\caption{Degree of linear polarization predicted by our HD~142527 model to be observed with VLT-NACO. The values compare very well with those obtained by \citet{2014ApJ...781...87A}. The field of view of the image is 2.85 arcsec.}
\label{fig:NIRpol}
\end{figure}

Though we did not use imaging data in our fitting procedure, another constraint on  the scattered light of the HD~142527 outer disk comes from polarimetric imaging using NACO at the VLT \citep{2014ApJ...781...87A}. We produce a synthetic image, which we smooth using a point spread function typical for the VLT with active adaptive optics. The resulting degree of polarization image is shown in Fig.~\ref{fig:NIRpol}. We find that in our model image the asymmetry from the observations and the overall degree of polarization are reproduced very well \citep[compare to][]{2014ApJ...781...87A}. The inner disk is not visible in the polarimetric image since it is very small and its polarization signal integrated over the spatial extend of the point spread function is negligible.

The 3\,$\mu$m resonance of water ice, which is abundantly observed in extinction in molecular clouds, causes a dip in the scattering efficiency of moderately large grains.
\citet{2009ApJ...690L.110H} took images of the outer disk in different filters in the near infrared. Using filters in and outside of the 3$\,\mu$m ice band they found a depression around 3$\,\mu$m, typical for scattering off ice covered grains. Thus, there must be icy grains on the surface of the outer disk. To compare with these observations we create a spectrum of our model where we remove emission from the inner 0.5'', so we only simulate and detect flux from the outer disk. We indeed detect the scattered light ice feature very clearly as is shown in Fig.~\ref{fig:OuterDisk}. 

\section{Discussion}
\label{sec:discussion}

\subsection{Abundance of water ice}
\label{sec:ice abundance}

\begin{figure*}[!t]
\centerline{\resizebox{\hsize}{!}{\includegraphics{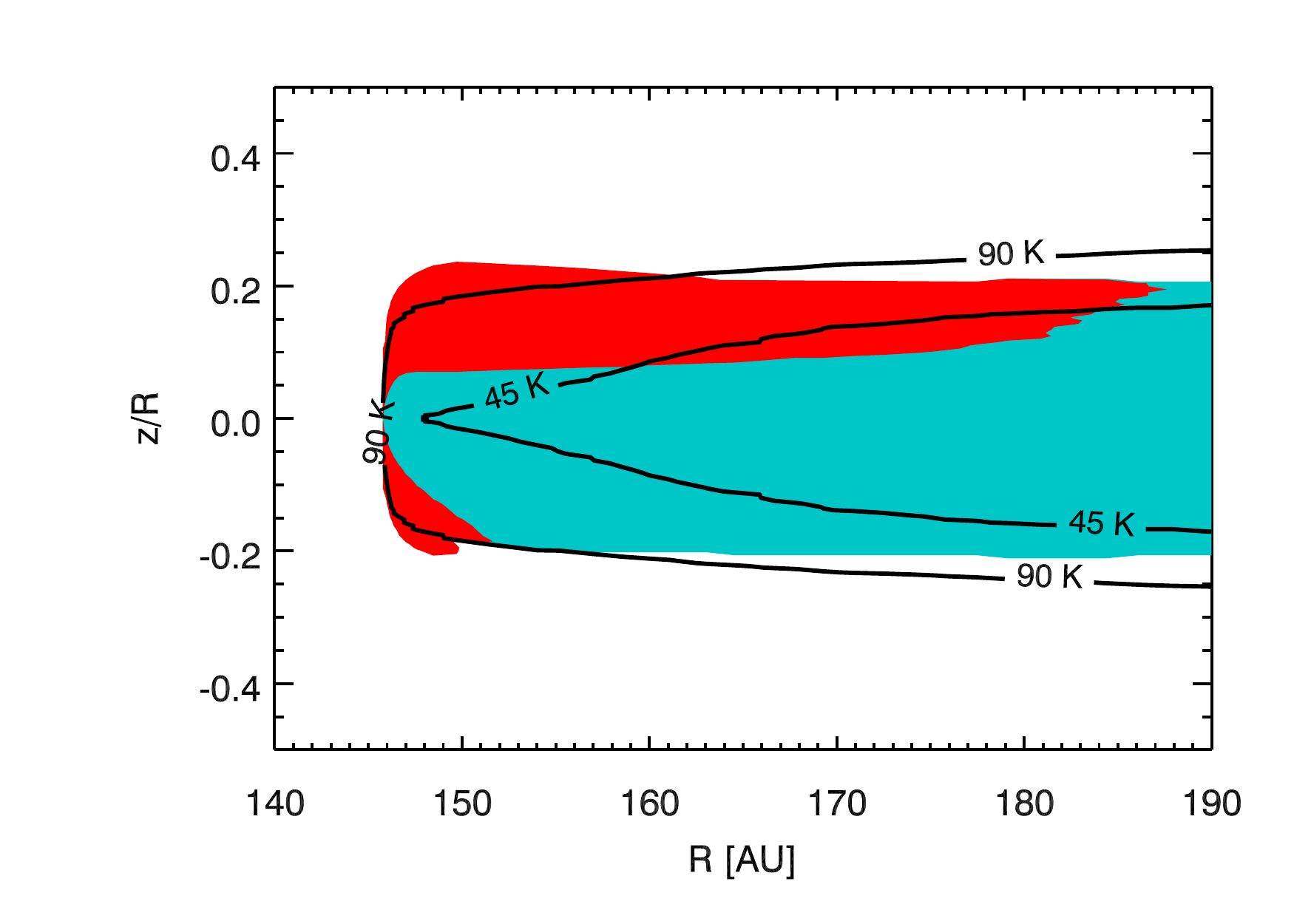}\includegraphics{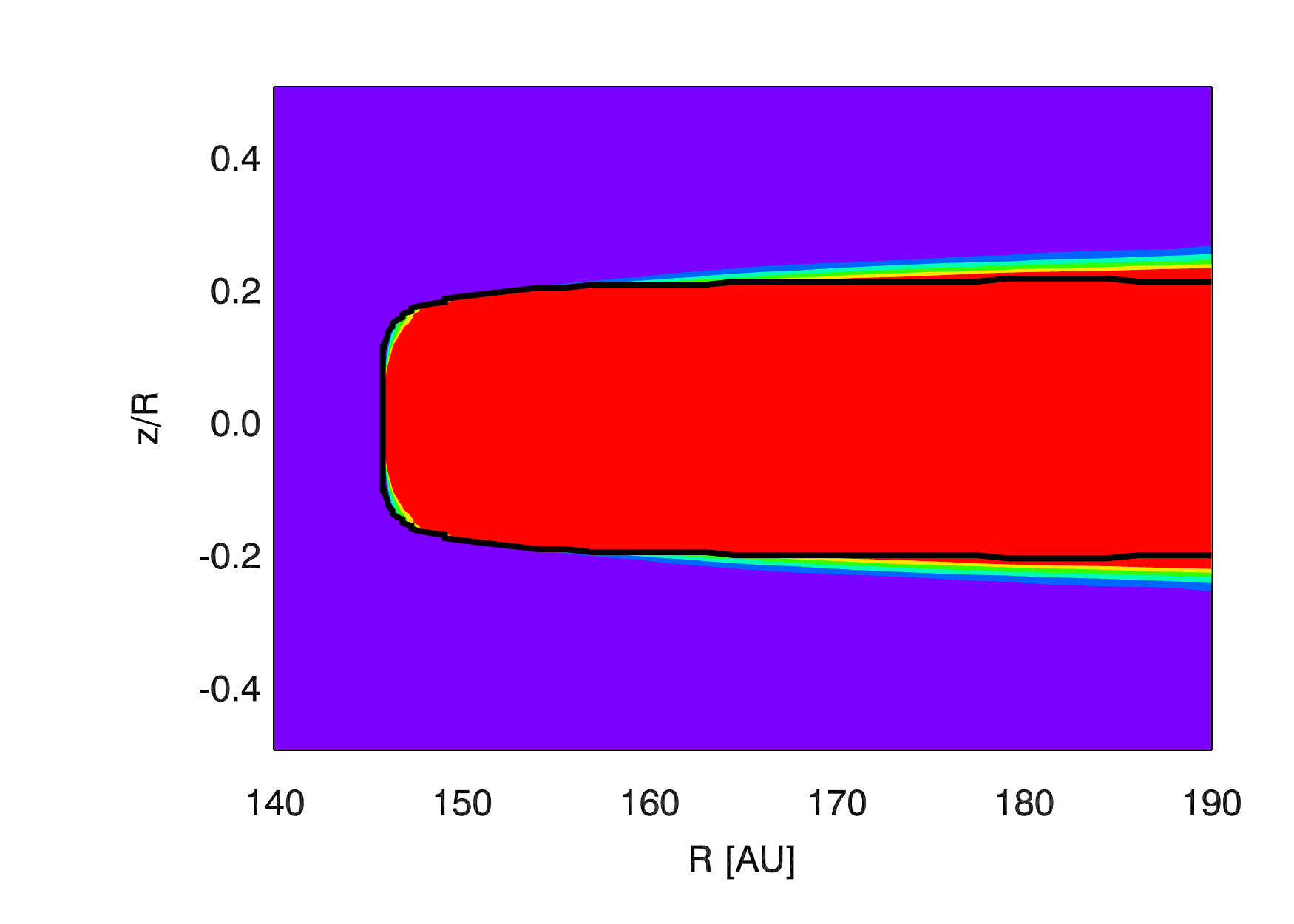}}}
\caption{\emph{Left panel:} The red area indicates the region in the disk where 90\% of the flux at 43\,$\mu$m originates. The blue area is the region where ice can exist, i.e. the temperature is low enough, the density high enough, and the UV field weak enough. The contours show the temperature of the ice according to the feature position (45\,K) and the crystallization temperature at one orbital timescale. \emph{Right panel:} Number of orbits required for the ice to crystallize. Colorscale: less than an orbit (purple), between 1 and 10 orbits (dark blue), between 10 and 100 orbits (light blue), between 100 and 1000 orbits (green), between 1000 and 10000 orbits (yellow), and above 10000 orbits (red). The thick black line indicates the region where ice can exist.}
\label{fig:contours}
\end{figure*}

We have presented the detection of water ice with ISO and Herschel in the protoplanetary disk surrounding HD~142527. The increase of the mass in solid material in the region where there is ice compared to where there is no ice has been estimated to range from a factor of 1.6 \citep[see e.g.][]{2011Icar..212..416M} to a factor of 4.2 \citep[see e.g.][]{2006plfo.book..129T}. Here we measure the strength of this jump directly from the ice to silicate ratio in the water ice emitting layer (i.e. indicated by the red region in Fig.~\ref{fig:contours}, explained in the next section in more detail) which we determine to be 1.6$^{+0.9}_{-0.6}$ (see Table~\ref{tab:parameters}). This value is surprisingly close to the solar system mixture value derived by \cite{2011Icar..212..416M} based on solar abundance arguments. From the abundances of all dust and ice species in our model fit, we can compute the elemental composition and compare this to the Solar composition. In Table~\ref{tab:abundances} we compute the fraction of all elements locked up in the different components of the model fit. Here we assume that all Si is locked in the silicates. Compared with Solar abundances from \citet{2009ARA&A..47..481A} all the oxygen atoms are used up by the solid components. The slight shortage of oxygen atoms in our model falls easily within the error margins. The number of Fe and S atoms remaining is approximately equal, consistent with the findings in our Solar system that sulfur is usually locked up in solid FeS (a dust component not considered in our analysis). We are however, left with a significant Mg abundance, which indicates that the silicate composition as used in our model is likely not complete and should be significantly more Mg rich. Note that the compositional analysis of the silicates by \citet{2005A&A...437..189V}, from which we took our silicate mixture, did not do a detailed analysis of the amorphous component in terms of Fe/Mg ratio. An attempt to get a handle on the Mg content of the amorphous silicates in disks was done by \citet{2010ApJ...721..431J} who find that Mg rich silicates provide a better match to the Spitzer spectra. The conclusion is that the mixture we find is consistent with elemental abundance constraints and that roughly 77$^{+10}_{-7}$\,\% of the available oxygen atoms are locked up in water ice.

Here we find an ice/silicate ratio that accounts for all the oxygen, and thus is the maximum ice reservoir one can imagine. In previous studies, the abundance of ice has been estimated to be significantly below that, leading to speculations of a vertically varying ice abundance due to enhanced grain growth from icy grains and subsequent settling of these grains below the visible surface of the disk. 
For example, our conclusion differs from the estimates of the ice abundance by \citet{2015ApJ...799..162M}, who estimated that only half of the predicted ice was accounted for in the spectra of T Tauri disks. A careful reanalysis of the low signal-to-noise features used in that study seems justified in the light of our ice abundance estimates to examine the significance of the difference.
Another possibility is that this difference is somehow related to the complex, dynamical geometry of the system, which should be a topic of future study.

\begin{table}[!tb]
\caption{The elemental composition of the mixture normalized to Si.}
\begin{center}
\begin{tabular}{l|c|c|c|c|c|c}
name			&	Si	&	Mg	&	Fe	&	S	&	C			&	O	 \\
\hline
Silicates			&	1.0	&	0.65	&	0.35	&	-	&	-			&	3.0	\\
Carbon			&	-	&	-	&	-	&	-	&	1.9 $^{\pm 0.9}$	&	-	\\
Water ice			&	-	&	-	&	-	&	-	&	-			&	9.9 $^{+5.6}_{-3.7}$	\\
\hline
Total				&	1.0	&	0.65	&	0.35	&	0.00	&	1.9 $^{\pm 0.9}$&	12.9	$^{+5.6}_{-3.7}$\\
Solar	\footnotemark	&	1.0	&	1.23	&	0.98	&	0.41	&	8.3			&	15.1	\\
\hline
Remaining		&	0.0	&	0.58	&	0.63	&	0.41	&	6.4 $^{\pm 0.9}$&	2.2	$^{+3.7}_{-5.6}$\\
\end{tabular}
\end{center}
\label{tab:abundances}
\end{table}
\footnotetext{Solar abundances taken from \citep{2009ARA&A..47..481A}}

\subsection{Origin of the crystallinity of the ice}

We have shown that the surface layer of the outer disk of HD~142527 contains a large abundance of crystalline water ice. Crystalline water ice forms in the disk at temperatures above $\sim90\,$K, warmer than the temperature where we detect it, which is around $45\,$K. We derive an upper limit of 40\,\% of the ice to be in amorphous form (see Table~\ref{tab:mineralogy}), but the modeling fit is also consistent with 100\,\% crystalline ice. 

\subsection{Crystallization and origin of ice}

The conditions in a laboratory on Earth are different from the conditions in a protoplanetary disk. Especially the available timescales are significantly longer in a protoplanetary disk. Therefore, we have to carefully consider what happens to the ice when it stays at a certain temperature for a longer period of time. In the laboratory a crystallization temperature of 110\,K is typically obtained for a crystallization timescale on the order of 1 day. The temperature needed for crystallization decreases if one allows more time for the transition to occur. Using the equation for the temperature dependent crystal growth rate from \cite{1996ApJ...473.1104J} the timescale for crystallization scales as,
\begin{equation}
\label{eq:timescale}
t_\mathrm{disk} = t_\mathrm{lab} \exp\left(-(E_a/R)\left\{\frac{1}{T_\mathrm{lab}}-\frac{1}{T_\mathrm{disk}}\right\}\right).
\end{equation}
Here $t_\mathrm{disk}$ and $t_\mathrm{lab}$ are the crystallization timescale in the disk and in the laboratory respectively, $T_\mathrm{disk}$ and $T_\mathrm{lab}$ are the crystallization temperatures in the disk and the laboratory, $E_a$ is the crystallization activation energy \citep[taken to be $E_a=59.2\,$kJ/mol after][]{2009AGUFM.P23E..03B}, and $R$ is the gas constant ($R=8.314\cdot10^{-3}\,$kJ/mol/K). For the disk we have to insert the local dynamical timescale, since this is the typical timescale that a dust grain can spend at a given location. Most dynamical processes act on a timescale on the order of the orbital timescale. At the inner edge of the outer wall (a radius of 130\,AU) this timescale is typically 1000\,yr, corresponding to a crystallization temperature of $T_\mathrm{disk}=91\,$K instead of the 110\,K measured in the laboratory.

In order to be able to plot some key features of the ice and temperature structure in the disk, we remove the inner disk to obtain a circularly symmetric model of the outer disk that we can map onto a 2D figure. Removing the inner disk implies that we are basically considering here the non-shadowed parts of the outer disk alone. In the left panel of Fig.~\ref{fig:contours} we plot the region in the disk where the 43\,$\mu$m flux originates. The red region indicates where 90\% of the flux at 43\,$\mu$m comes from. The blue region indicates the region where in the outer disk ice can exist according to Eq.~\ref{eq:finalTsub}. Also drawn are contours for 45\,K (the temperature of the ice according to the peak position) and 90\,K (the temperature where ice can crystallize in one orbital timescale). In the right panel we plot the number of orbits required for ice to crystallize at that location in the disk according to Eq.~\ref{eq:timescale} (colorscale as indicated in the caption). Also plotted in that panel by the thick black line is the location where ice can exist. As can be seen from these figures, by far most of the ice is in a region where it is expected to be amorphous. Also, most of the flux at 43\,$\mu$m originates in this region. 
We now consider a few options for generating the amount of crystalline ice seen in the outer disk.

\emph{Condensation in the gap:}
At a distance of $\sim100\,$AU from the star, the temperature of the dust grains in the optically thin limit reaches 100\,K, sufficient to make crystalline water ice. However, this is right inside the gap. The computations performed with ProDiMo confirmed that the vapor pressure of H$_2$O here is too low to form water ice (see also Fig.~\ref{fig:contours}).

\emph{In situ formation in the outer disk:}
In the outer disk, ice condensation can take place. As mentioned before, at the equilibrium temperature at that location the ice will form amorphous. In Fig.~\ref{fig:contours} it can be seen that there are small regions where the ice could crystallize in 10 to 100 orbits. However, these are regions high up in the disk, where the densities are low, so there is not much ice there. Crystallizing the entire region responsible for the 43\,$\mu$m feature requires substantial mixing in order to process the entire material in a column through the small crystallization region. This basically implies that we have to divide the timescale required for crystallization by the fraction of the ice that actually is at high enough temperatures to compute the timescale for full crystallization. This results in a timescale significantly longer than the lifetime of the disk. The crystalline ice we detect therefore requires (transient) heating above 100\,K (where the ice crystallizes almost instantaneously). These events have to heat the material somewhere between 100 and 160\,K such that the ice crystallizes without evaporating. Alternatively, the heating can be to temperatures above 160\,K, where the ice evaporates, and the cooling has to be sufficiently slow to allow the ice to reform crystalline. If the source of the heating is a transient accretion burst, we can compute the temperature structure using the method of \citet{2011Icar..212..416M}. We find that an accretion burst of 10$^{-4}\,$M$_{\sun}$/yr, which is very high, provides enough energy (both locally produced and through irradiation from the inner accretion shock) to crystallize the ice.

\emph{Heating by a companion inside the gap:}
One of the possible causes of the gap is the presence of a companion close to the inner edge of the outer disk. The luminosity of this companion can heat up the dust grains in the wall of the outer disk when it passes. If we assume that the companion is one Hill radius away from the outer edge of the gap, which is at least around 10\,AU at this distance from the star, the required luminosity of the companion to heat the grains from 80 to 100\,K is around $8\cdot 10^{-3}\,$L$_{\star}$. This is almost two orders of magnitude above the detection limits obtained by \citet{2012A&A...546A..24R}, ruling out this scenario.

\emph{Transportation from the inner disk:} 
In the midplane of the inner disk, the temperature and densities are right for creating crystalline water ice. At the outer edge of the inner disk, around 30 - 40\,AU, the temperature at the surface of the disk is low enough for the ice to survive for a moderate amount of time. One could imagine the crystalline ice being created there and transported through the gap by the radiation pressure from the star \citep[similar to the mechanism proposed by][]{2009Natur.459..227V}. Normally in a protoplanetary disk, radiation pressure cannot move the grains very far because they are stopped by the gas. However, in this case transportation through the low density region, the gap, seems possible depending on the currently unknown gas density inside the gap.

\emph{Creation in the gap by collisions:} Inside a larger body, the water ice can be preserved and protected against evaporation, photo destruction, and amorphization. Indeed, the ice in Kuiper Belt objects can be crystalline \citep{2004Natur.432..731J}, even at the surface. The gap in the disk is very large, and deep imaging has so far only revealed upper limits to possible planetary companions inside the gap \citep{2012A&A...546A..24R}. If this very large gap is created by multiple planets, it can very well be that it also houses a collection of large planetesimals. When these bodies collide, they can free their crystalline interior in the form of small dust grains, which are pushed towards the outer disk by radiation pressure or a stellar wind. This mechanism was also proposed to explain the presence of crystalline forsterite particles at the inner edge of the outer disk in the HD100546 system \citep{2003A&A...401..577B, 2011A&A...531A..93M} and the crystalline water ice in T Tauri disks \citep{2015ApJ...799..162M}. The 0.5 Earth masses of water ice corresponds to roughly $5\cdot10^{7}$ one-km-sized planetesimals.

Whatever the formation mechanism, the detection of crystalline water ice is consistent with observations on solar system satellites, where most are detected to have crystalline ice at least at the surface \citep{2013sssi.book..371M}, also for satellites that seem too cold to have crystalline ice.

\section{Conclusions}

We report here on the far-infrared spectrum of the Herbig star HD~142527. The Herschel PACS spectrum of this source does not confirm the previous claim of a 105\,$\mu$m feature in the ISO LWS spectrum. We identify calibration issues with the ISO LWS spectrum as the source of the confusion. The $62\,\mu$m water ice feature is clearly detected in the PACS spectrum, as is the 43\,$\mu$m water ice feature in the ISO SWS spectrum.

The full SED of the source is modeled in detail, with a special focus on the detection of crystalline water ice.  Assuming a homogeneous composition of the outer disk, and ice condensation where possible, we infer an ice/silicate ratio of 1.6~$^{+0.9}_{-0.6}$. This implies that roughly 80\,\% of the available oxygen atoms are locked up in water ice. We compute the elemental abundances of the derived dust/ice mixture and conclude that it is consistent with Solar abundance constraints. Furthermore, from our modelling we find that the scaleheight of the disk in the inner regions is significantly above hydrostatic equilibrium, while in the outer parts of the disk it is close to hydrostatic equilibrium. This is including detailed settling with a very low value of the turbulent mixing strength, $\alpha$.

We identify crystalline water ice in a location of the disk where equilibrium temperatures are too low to form crystalline ice in situ. In addition, the spectral position of the 43\,$\mu$m ice feature is indicative of ice at a temperature of $45$\,K.  At these temperatures amorphous ice is expected to form. We discuss possible scenarios for forming or transporting the crystalline ice there. We conclude that a plausible scenario is formation inside the large 146\,AU gap through collisions of icy planetesimals. Another possibility is crystallization trough extreme accretion events liberating enough energy to heat up and crystallize the ice in the outer regions.

Our findings of a large region in the disk where the solid mass is dominated by water ice is promising for planet formation models that require water ice as a mechanism for increasing the dust mass and sticking probability. It also provides a huge reservoir of volatile elements that may contribute at a later stage to the enrichment of dry planets closer to the star.

\begin{acknowledgements}
We would like to thank Inga Kamp for valuable discussions on the influence of UV radiation.
Support for this work, part of the Herschel Open Time Key
Project Program, was provided by NASA through an award
issued by the Jet Propulsion Laboratory, California Institute of
Technology.
M.M. acknowledges funding from the EU FP7- 2011 under Grant Agreement No 284405.
OSIA is a joint development of the SWS consortium. Contributing
institutes are SRON, MPE, KUL and the ESA Astrophysics Division. The
ISO Spectral Analysis Package (ISAP) is a joint development by the LWS
and SWS Instrument Teams and Data Centers. Contributing institutes are
CESR, IAS, IPAC, MPE, RAL and SRON.
\end{acknowledgements}

\bibliographystyle{aa}
\bibliography{biblio}

\begin{thebibliography}{67}
\expandafter\ifx\csname natexlab\endcsname\relax\def\natexlab#1{#1}\fi

\bibitem[{{Asplund} {et~al.}(2009){Asplund}, {Grevesse}, {Sauval}, \&
  {Scott}}]{2009ARA&A..47..481A}
{Asplund}, M., {Grevesse}, N., {Sauval}, A.~J., \& {Scott}, P. 2009, \araa, 47,
  481

\bibitem[{{Avenhaus} {et~al.}(2014){Avenhaus}, {Quanz}, {Schmid}, {Meyer},
  {Garufi}, {Wolf}, \& {Dominik}}]{2014ApJ...781...87A}
{Avenhaus}, H., {Quanz}, S.~P., {Schmid}, H.~M., {et~al.} 2014, \apj, 781, 87

\bibitem[{{Bans} \& {K{\"o}nigl}(2012)}]{2012ApJ...758..100B}
{Bans}, A. \& {K{\"o}nigl}, A. 2012, \apj, 758, 100

\bibitem[{{Baragiola} {et~al.}(2009){Baragiola}, {Burke}, \&
  {Fama}}]{2009AGUFM.P23E..03B}
{Baragiola}, R.~A., {Burke}, D.~J., \& {Fama}, M.~A. 2009, AGU Fall Meeting
  Abstracts

\bibitem[{{Biller} {et~al.}(2012){Biller}, {Lacour}, {Juh{\'a}sz}, {Benisty},
  {Chauvin}, {Olofsson}, {Pott}, {M{\"u}ller}, {Sicilia-Aguilar}, {Bonnefoy},
  {Tuthill}, {Thebault}, {Henning}, \& {Crida}}]{2012ApJ...753L..38B}
{Biller}, B., {Lacour}, S., {Juh{\'a}sz}, A., {et~al.} 2012, \apjl, 753, L38

\bibitem[{{Birnstiel} {et~al.}(2013){Birnstiel}, {Dullemond}, \&
  {Pinilla}}]{2013A&A...550L...8B}
{Birnstiel}, T., {Dullemond}, C.~P., \& {Pinilla}, P. 2013, \aap, 550, L8

\bibitem[{{Bouwman} {et~al.}(2003){Bouwman}, {de Koter}, {Dominik}, \&
  {Waters}}]{2003A&A...401..577B}
{Bouwman}, J., {de Koter}, A., {Dominik}, C., \& {Waters}, L.~B.~F.~M. 2003,
  \aap, 401, 577

\bibitem[{{Canovas} {et~al.}(2013){Canovas}, {M{\'e}nard}, {Hales},
  {Jord{\'a}n}, {Schreiber}, {Casassus}, {Gledhill}, \&
  {Pinte}}]{2013A&A...556A.123C}
{Canovas}, H., {M{\'e}nard}, F., {Hales}, A., {et~al.} 2013, \aap, 556, A123

\bibitem[{{Casassus} {et~al.}(2013){Casassus}, {van der Plas}, {M}, {Dent},
  {Fomalont}, {Hagelberg}, {Hales}, {Jord{\'a}n}, {Mawet}, {M{\'e}nard},
  {Wootten}, {Wilner}, {Hughes}, {Schreiber}, {Girard}, {Ercolano}, {Canovas},
  {Rom{\'a}n}, \& {Salinas}}]{2013Natur.493..191C}
{Casassus}, S., {van der Plas}, G., {M}, S.~P., {et~al.} 2013, \nat, 493, 191

\bibitem[{{Charbonneau}(1995)}]{1995ApJS..101..309C}
{Charbonneau}, P. 1995, \apjs, 101, 309

\bibitem[{{Chiang} {et~al.}(2001){Chiang}, {Joung}, {Creech-Eakman}, {Qi},
  {Kessler}, {Blake}, \& {van Dishoeck}}]{2001ApJ...547.1077C}
{Chiang}, E.~I., {Joung}, M.~K., {Creech-Eakman}, M.~J., {et~al.} 2001, \apj,
  547, 1077

\bibitem[{{Chiavassa} {et~al.}(2005){Chiavassa}, {Ceccarelli}, {Tielens},
  {Caux}, \& {Maret}}]{2005A&A...432..547C}
{Chiavassa}, A., {Ceccarelli}, C., {Tielens}, A.~G.~G.~M., {Caux}, E., \&
  {Maret}, S. 2005, \aap, 432, 547

\bibitem[{{Chihara} {et~al.}(2001){Chihara}, {Koike}, \&
  {Tsuchiyama}}]{2001PASJ...53..243C}
{Chihara}, H., {Koike}, C., \& {Tsuchiyama}, A. 2001, \pasj, 53, 243

\bibitem[{{Christiaens} {et~al.}(2014){Christiaens}, {Casassus}, {Perez}, {van
  der Plas}, \& {M{\'e}nard}}]{2014ApJ...785L..12C}
{Christiaens}, V., {Casassus}, S., {Perez}, S., {van der Plas}, G., \&
  {M{\'e}nard}, F. 2014, \apjl, 785, L12

\bibitem[{{Clegg} {et~al.}(1996){Clegg}, {Ade}, {Armand}, {Baluteau}, {Barlow},
  {Buckley}, {Berges}, {Burgdorf}, {Caux}, {Ceccarelli}, {Cerulli}, {Church},
  {Cotin}, {Cox}, {Cruvellier}, {Culhane}, {Davis}, {di Giorgio}, {Diplock},
  {Drummond}, {Emery}, {Ewart}, {Fischer}, {Furniss}, {Glencross},
  {Greenhouse}, {Griffin}, {Gry}, {Harwood}, {Hazell}, {Joubert}, {King},
  {Lim}, {Liseau}, {Long}, {Lorenzetti}, {Molinari}, {Murray}, {Naylor},
  {Nisini}, {Norman}, {Omont}, {Orfei}, {Patrick}, {Pequignot}, {Pouliquen},
  {Price}, {Nguyen-Q-Rieu}, {Rogers}, {Robinson}, {Saisse}, {Saraceno},
  {Serra}, {Sidher}, {Smith}, {Smith}, {Spinoglio}, {Swinyard}, {Texier},
  {Towlson}, {Trams}, {Unger}, \& {White}}]{1996A&A...315L..38C}
{Clegg}, P.~E., {Ade}, P.~A.~R., {Armand}, C., {et~al.} 1996, \aap, 315, L38

\bibitem[{{Close} {et~al.}(2014){Close}, {Follette}, {Males}, {Puglisi},
  {Xompero}, {Apai}, {Najita}, {Weinberger}, {Morzinski}, {Rodigas}, {Hinz},
  {Bailey}, \& {Briguglio}}]{2014ApJ...781L..30C}
{Close}, L.~M., {Follette}, K.~B., {Males}, J.~R., {et~al.} 2014, \apjl, 781,
  L30

\bibitem[{{Creech-Eakman} {et~al.}(2002){Creech-Eakman}, {Chiang}, {Joung},
  {Blake}, \& {van Dishoeck}}]{2002A&A...385..546C}
{Creech-Eakman}, M.~J., {Chiang}, E.~I., {Joung}, R.~M.~K., {Blake}, G.~A., \&
  {van Dishoeck}, E.~F. 2002, \aap, 385, 546

\bibitem[{{de Graauw} {et~al.}(1996){de Graauw}, {Haser}, {Beintema},
  {Roelfsema}, {van Agthoven}, {Barl}, {Bauer}, {Bekenkamp}, {Boonstra},
  {Boxhoorn}, {Cote}, {de Groene}, {van Dijkhuizen}, {Drapatz}, {Evers},
  {Feuchtgruber}, {Frericks}, {Genzel}, {Haerendel}, {Heras}, {van der Hucht},
  {van der Hulst}, {Huygen}, {Jacobs}, {Jakob}, {Kamperman}, {Katterloher},
  {Kester}, {Kunze}, {Kussendrager}, {Lahuis}, {Lamers}, {Leech}, {van der
  Lei}, {van der Linden}, {Luinge}, {Lutz}, {Melzner}, {Morris}, {van Nguyen},
  {Ploeger}, {Price}, {Salama}, {Schaeidt}, {Sijm}, {Smoorenburg}, {Spakman},
  {Spoon}, {Steinmayer}, {Stoecker}, {Valentijn}, {Vandenbussche}, {Visser},
  {Waelkens}, {Waters}, {Wensink}, {Wesselius}, {Wiezorrek}, {Wieprecht},
  {Wijnbergen}, {Wildeman}, \& {Young}}]{1996A&A...315L..49D}
{de Graauw}, T., {Haser}, L.~N., {Beintema}, D.~A., {et~al.} 1996, \aap, 315,
  L49

\bibitem[{{Dorschner} {et~al.}(1995){Dorschner}, {Begemann}, {Henning},
  {J\"ager}, \& {Mutschke}}]{1995A&A...300..503D}
{Dorschner}, J., {Begemann}, B., {Henning}, T., {J\"ager}, C., \& {Mutschke},
  H. 1995, \aap, 300, 503

\bibitem[{{Dubrulle} {et~al.}(1995){Dubrulle}, {Morfill}, \&
  {Sterzik}}]{1995Icar..114..237D}
{Dubrulle}, B., {Morfill}, G., \& {Sterzik}, M. 1995, \icarus, 114, 237

\bibitem[{{Fabian} {et~al.}(2001){Fabian}, {Henning}, {J{\"a}ger}, {Mutschke},
  {Dorschner}, \& {Wehrhan}}]{2001A&A...378..228F}
{Fabian}, D., {Henning}, T., {J{\"a}ger}, C., {et~al.} 2001, \aap, 378, 228

\bibitem[{{Ferrarotti} {et~al.}(2000){Ferrarotti}, {Gail}, {Degiorgi}, \&
  {Ott}}]{2000A&A...357L..13F}
{Ferrarotti}, A., {Gail}, H.-P., {Degiorgi}, L., \& {Ott}, H.~R. 2000, \aap,
  357, L13

\bibitem[{{Green} {et~al.}(2016){Green}, {Yang}, {Evans}, {Karska}, {Herczeg},
  {van Dishoeck}, {Lee}, {Larson}, \& {Bouwman}}]{2016AJ....151...75G}
{Green}, J.~D., {Yang}, Y.-L., {Evans}, II, N.~J., {et~al.} 2016, \aj, 151, 75

\bibitem[{{Habing}(1968)}]{1968BAN....19..421H}
{Habing}, H.~J. 1968, \bain, 19, 421

\bibitem[{{Hartmann} {et~al.}(1998){Hartmann}, {Calvet}, {Gullbring}, \&
  {D'Alessio}}]{1998ApJ...495..385H}
{Hartmann}, L., {Calvet}, N., {Gullbring}, E., \& {D'Alessio}, P. 1998, \apj,
  495, 385

\bibitem[{{Honda} {et~al.}(2009){Honda}, {Inoue}, {Fukagawa}, {Oka},
  {Nakamoto}, {Ishii}, {Terada}, {Takato}, {Kawakita}, {Okamoto}, {Shibai},
  {Tamura}, {Kudo}, \& {Itoh}}]{2009ApJ...690L.110H}
{Honda}, M., {Inoue}, A.~K., {Fukagawa}, M., {et~al.} 2009, \apjl, 690, L110

\bibitem[{{Hughes} {et~al.}(2008){Hughes}, {Wilner}, {Qi}, \&
  {Hogerheijde}}]{2008ApJ...678.1119H}
{Hughes}, A.~M., {Wilner}, D.~J., {Qi}, C., \& {Hogerheijde}, M.~R. 2008, \apj,
  678, 1119

\bibitem[{{J\"ager} {et~al.}(1998){J\"ager}, {Molster}, {Dorschner}, {Henning},
  {Mutschke}, \& {Waters}}]{1998A&A...339..904J}
{J\"ager}, C., {Molster}, F.~J., {Dorschner}, J., {et~al.} 1998, \aap, 339, 904

\bibitem[{{Jenniskens} \& {Blake}(1996)}]{1996ApJ...473.1104J}
{Jenniskens}, P. \& {Blake}, D.~F. 1996, \apj, 473, 1104

\bibitem[{{Jewitt} \& {Luu}(2004)}]{2004Natur.432..731J}
{Jewitt}, D.~C. \& {Luu}, J. 2004, \nat, 432, 731

\bibitem[{{Juh{\'a}sz} {et~al.}(2010){Juh{\'a}sz}, {Bouwman}, {Henning},
  {Acke}, {van den Ancker}, {Meeus}, {Dominik}, {Min}, {Tielens}, \&
  {Waters}}]{2010ApJ...721..431J}
{Juh{\'a}sz}, A., {Bouwman}, J., {Henning}, T., {et~al.} 2010, \apj, 721, 431

\bibitem[{{Kama} {et~al.}(2009){Kama}, {Min}, \&
  {Dominik}}]{2009A&A...506.1199K}
{Kama}, M., {Min}, M., \& {Dominik}, C. 2009, \aap, 506, 1199

\bibitem[{{Kessler} {et~al.}(1996){Kessler}, {Steinz}, {Anderegg}, {Clavel},
  {Drechsel}, {Estaria}, {Faelker}, {Riedinger}, {Robson}, {Taylor}, \&
  {Xim{\'e}nez de Ferr{\'a}n}}]{1996A&A...315L..27K}
{Kessler}, M.~F., {Steinz}, J.~A., {Anderegg}, M.~E., {et~al.} 1996, \aap, 315,
  L27

\bibitem[{{Klahr} \& {Henning}(1997)}]{1997Icar..128..213K}
{Klahr}, H.~H. \& {Henning}, T. 1997, \icarus, 128, 213

\bibitem[{{Koike} \& {Shibai}(1990)}]{1990MNRAS.246..332K}
{Koike}, C. \& {Shibai}, H. 1990, \mnras, 246, 332

\bibitem[{{Lacour} {et~al.}(2015){Lacour}, {Biller}, {Cheetham}, {Greenbaum},
  {Pearce}, {Marino}, {Tuthill}, {Pueyo}, {Mamajek}, {Girard},
  {Sivaramakrishnan}, {Bonnefoy}, {Baraffe}, {Chauvin}, {Olofsson}, {Juhasz},
  {Benisty}, {Pott}, {Sicilia-Aguilar}, {Henning}, {Cardwell}, {Goodsell},
  {Graham}, {Hibon}, {Ingraham}, {Konopacky}, {Macintosh}, {Oppenheimer},
  {Perrin}, {Rantakyr{\"o}}, {Sadakuni}, \& {Thomas}}]{2015arXiv151109390L}
{Lacour}, S., {Biller}, B., {Cheetham}, A., {et~al.} 2015, ArXiv e-prints

\bibitem[{{Maaskant} {et~al.}(2013){Maaskant}, {Honda}, {Waters}, {Tielens},
  {Dominik}, {Min}, {Verhoeff}, {Meeus}, \& {van den
  Ancker}}]{2013A&A...555A..64M}
{Maaskant}, K.~M., {Honda}, M., {Waters}, L.~B.~F.~M., {et~al.} 2013, \aap,
  555, A64

\bibitem[{{Malfait} {et~al.}(1999){Malfait}, {Waelkens}, {Bouwman}, {de Koter},
  \& {Waters}}]{1999A&A...345..181M}
{Malfait}, K., {Waelkens}, C., {Bouwman}, J., {de Koter}, A., \& {Waters},
  L.~B.~F.~M. 1999, \aap, 345, 181

\bibitem[{{Marino} {et~al.}(2015){Marino}, {Perez}, \&
  {Casassus}}]{2015ApJ...798L..44M}
{Marino}, S., {Perez}, S., \& {Casassus}, S. 2015, \apjl, 798, L44

\bibitem[{{Mastrapa} {et~al.}(2013){Mastrapa}, {Grundy}, \&
  {Gudipati}}]{2013sssi.book..371M}
{Mastrapa}, R.~M.~E., {Grundy}, W.~M., \& {Gudipati}, M.~S. 2013, {Amorphous
  and Crystalline H$_{2}$O-Ice}, ed. M.~S. {Gudipati} \& J.~{Castillo-Rogez},
  371

\bibitem[{{McClure} {et~al.}(2015){McClure}, {Espaillat}, {Calvet}, {Bergin},
  {D'Alessio}, {Watson}, {Manoj}, {Sargent}, \&
  {Cleeves}}]{2015ApJ...799..162M}
{McClure}, M.~K., {Espaillat}, C., {Calvet}, N., {et~al.} 2015, \apj, 799, 162

\bibitem[{{McClure} {et~al.}(2012){McClure}, {Manoj}, {Calvet}, {Adame},
  {Espaillat}, {Watson}, {Sargent}, {Forrest}, \&
  {D'Alessio}}]{2012ApJ...759L..10M}
{McClure}, M.~K., {Manoj}, P., {Calvet}, N., {et~al.} 2012, \apjl, 759, L10

\bibitem[{{Min} {et~al.}(2009){Min}, {Dullemond}, {Dominik}, {de Koter}, \&
  {Hovenier}}]{2009A&A...497..155M}
{Min}, M., {Dullemond}, C.~P., {Dominik}, C., {de Koter}, A., \& {Hovenier},
  J.~W. 2009, \aap, 497, 155

\bibitem[{{Min} {et~al.}(2011){Min}, {Dullemond}, {Kama}, \&
  {Dominik}}]{2011Icar..212..416M}
{Min}, M., {Dullemond}, C.~P., {Kama}, M., \& {Dominik}, C. 2011, \icarus, 212,
  416

\bibitem[{{Min} {et~al.}(2005){Min}, {Hovenier}, \& {de
  Koter}}]{2005A&A...432..909M}
{Min}, M., {Hovenier}, J.~W., \& {de Koter}, A. 2005, \aap, 432, 909

\bibitem[{{Min} {et~al.}(2007){Min}, {Waters}, {de Koter}, {Hovenier},
  {Keller}, \& {Markwick-Kemper}}]{2007A&A...462..667M}
{Min}, M., {Waters}, L.~B.~F.~M., {de Koter}, A., {et~al.} 2007, \aap, 462, 667

\bibitem[{{Mulders} {et~al.}(2011){Mulders}, {Waters}, {Dominik}, {Sturm},
  {Bouwman}, {Min}, {Verhoeff}, {Acke}, {Augereau}, {Evans}, {Henning},
  {Meeus}, \& {Olofsson}}]{2011A&A...531A..93M}
{Mulders}, G.~D., {Waters}, L.~B.~F.~M., {Dominik}, C., {et~al.} 2011, \aap,
  531, A93

\bibitem[{{M{\"u}ller} {et~al.}(2014){M{\"u}ller}, {Balog}, {Nielbock}, {Lim},
  {Teyssier}, {Olberg}, {Klaas}, {Linz}, {Altieri}, {Pearson}, {Bendo}, \&
  {Vilenius}}]{Mueller2014}
{M{\"u}ller}, T., {Balog}, Z., {Nielbock}, M., {et~al.} 2014, Experimental
  Astronomy, 37, 253

\bibitem[{{Mutschke} {et~al.}(2008){Mutschke}, {Zeidler}, {Posch},
  {Kerschbaum}, {Baier}, \& {Henning}}]{2008A&A...492..117M}
{Mutschke}, H., {Zeidler}, S., {Posch}, T., {et~al.} 2008, \aap, 492, 117

\bibitem[{{Ott}(2010)}]{Ott2010}
{Ott}, S. 2010, in Astronomical Society of the Pacific Conference Series, Vol.
  434, Astronomical Data Analysis Software and Systems XIX, ed. {Y.~Mizumoto,
  K.-I.~Morita, \& M.~Ohishi}, 139

\bibitem[{{Poglitsch} {et~al.}(2010){Poglitsch}, {Waelkens}, {Geis},
  {Feuchtgruber}, {Vandenbussche}, {Rodriguez}, {Krause}, {Renotte}, {van
  Hoof}, {Saraceno}, {Cepa}, {Kerschbaum}, {Agn{\`e}se}, {Ali}, {Altieri},
  {Andreani}, {Augueres}, {Balog}, {Barl}, {Bauer}, {Belbachir}, {Benedettini},
  {Billot}, {Boulade}, {Bischof}, {Blommaert}, {Callut}, {Cara}, {Cerulli},
  {Cesarsky}, {Contursi}, {Creten}, {De Meester}, {Doublier}, {Doumayrou},
  {Duband}, {Exter}, {Genzel}, {Gillis}, {Gr{\"o}zinger}, {Henning},
  {Herreros}, {Huygen}, {Inguscio}, {Jakob}, {Jamar}, {Jean}, {de Jong},
  {Katterloher}, {Kiss}, {Klaas}, {Lemke}, {Lutz}, {Madden}, {Marquet},
  {Martignac}, {Mazy}, {Merken}, {Montfort}, {Morbidelli}, {M{\"u}ller},
  {Nielbock}, {Okumura}, {Orfei}, {Ottensamer}, {Pezzuto}, {Popesso},
  {Putzeys}, {Regibo}, {Reveret}, {Royer}, {Sauvage}, {Schreiber}, {Stegmaier},
  {Schmitt}, {Schubert}, {Sturm}, {Thiel}, {Tofani}, {Vavrek}, {Wetzstein},
  {Wieprecht}, \& {Wiezorrek}}]{Poglitsch2010}
{Poglitsch}, A., {Waelkens}, C., {Geis}, N., {et~al.} 2010, \aap, 518, L2

\bibitem[{{Pollack} {et~al.}(1994){Pollack}, {Hollenbach}, {Beckwith},
  {Simonelli}, {Roush}, \& {Fong}}]{1994ApJ...421..615P}
{Pollack}, J.~B., {Hollenbach}, D., {Beckwith}, S., {et~al.} 1994, \apj, 421,
  615

\bibitem[{{Posch} {et~al.}(2007){Posch}, {Baier}, {Mutschke}, \&
  {Henning}}]{2007ApJ...668..993P}
{Posch}, T., {Baier}, A., {Mutschke}, H., \& {Henning}, T. 2007, \apj, 668, 993

\bibitem[{{Preibisch} {et~al.}(1993){Preibisch}, {Ossenkopf}, {Yorke}, \&
  {Henning}}]{1993A&A...279..577P}
{Preibisch}, T., {Ossenkopf}, V., {Yorke}, H.~W., \& {Henning}, T. 1993, \aap,
  279, 577

\bibitem[{{Rameau} {et~al.}(2012){Rameau}, {Chauvin}, {Lagrange},
  {Th{\'e}bault}, {Milli}, {Girard}, \& {Bonnefoy}}]{2012A&A...546A..24R}
{Rameau}, J., {Chauvin}, G., {Lagrange}, A.-M., {et~al.} 2012, \aap, 546, A24

\bibitem[{{Servoin} \& {Piriou}(1973)}]{1973PSSBR..55..677S}
{Servoin}, J.~L. \& {Piriou}, B. 1973, Physica Status Solidi B Basic Research,
  55, 677

\bibitem[{{Smith} {et~al.}(1994){Smith}, {Robinson}, {Hyland}, \&
  {Carpenter}}]{1994MNRAS.271..481S}
{Smith}, R.~G., {Robinson}, G., {Hyland}, A.~R., \& {Carpenter}, G.~L. 1994,
  \mnras, 271, 481

\bibitem[{{Spitzer} \& {Kleinman}(1960)}]{1960PhRv..121.1324S}
{Spitzer}, W.~G. \& {Kleinman}, D.~A. 1960, Physical Review, 121, 1324

\bibitem[{{Thommes} \& {Duncan}(2006)}]{2006plfo.book..129T}
{Thommes}, E.~W. \& {Duncan}, M.~J. 2006, {The accretion of giant-planet
  cores}, ed. H.~{Klahr} \& W.~{Brandner} (Cambridge University Press), 129

\bibitem[{{van Boekel} {et~al.}(2004){van Boekel}, {Min}, {Leinert}, {Waters},
  {Richichi}, {Chesneau}, {Dominik}, {Jaffe}, {Dutrey}, {Graser}, {Henning},
  {de Jong}, {K{\"o}hler}, {de Koter}, {Lopez}, {Malbet}, {Morel}, {Paresce},
  {Perrin}, {Preibisch}, {Przygodda}, {Sch{\"o}ller}, \&
  {Wittkowski}}]{2004Natur.432..479V}
{van Boekel}, R., {Min}, M., {Leinert}, C., {et~al.} 2004, \nat, 432, 479

\bibitem[{{van Boekel} {et~al.}(2005){van Boekel}, {Min}, {Waters}, {de Koter},
  {Dominik}, {van den Ancker}, \& {Bouwman}}]{2005A&A...437..189V}
{van Boekel}, R., {Min}, M., {Waters}, L.~B.~F.~M., {et~al.} 2005, \aap, 437,
  189

\bibitem[{{van der Plas} {et~al.}(2014){van der Plas}, {Casassus}, {Menard},
  {Perez}, {Thi}, {Pinte}, \& {Christiaens}}]{2014arXiv1407.1735V}
{van der Plas}, G., {Casassus}, S., {Menard}, F., {et~al.} 2014, ArXiv e-prints

\bibitem[{{van der Wiel} {et~al.}(2014){van der Wiel}, {Naylor}, {Kamp},
  {M{\'e}nard}, {Thi}, {Woitke}, {Olofsson}, {Pontoppidan}, {Di Francesco},
  {Glauser}, {Greaves}, \& {Ivison}}]{2014MNRAS.444.3911V}
{van der Wiel}, M.~H.~D., {Naylor}, D.~A., {Kamp}, I., {et~al.} 2014, \mnras,
  444, 3911

\bibitem[{{Verhoeff} {et~al.}(2011){Verhoeff}, {Min}, {Pantin}, {Waters},
  {Tielens}, {Honda}, {Fujiwara}, {Bouwman}, {van Boekel}, {Dougherty}, {de
  Koter}, {Dominik}, \& {Mulders}}]{2011A&A...528A..91V}
{Verhoeff}, A.~P., {Min}, M., {Pantin}, E., {et~al.} 2011, \aap, 528, A91

\bibitem[{{Vinkovi{\'c}}(2009)}]{2009Natur.459..227V}
{Vinkovi{\'c}}, D. 2009, \nat, 459, 227

\bibitem[{{Woitke} {et~al.}(2009){Woitke}, {Kamp}, \&
  {Thi}}]{2009A&A...501..383W}
{Woitke}, P., {Kamp}, I., \& {Thi}, W.-F. 2009, \aap, 501, 383

\bibitem[{{Woitke} {et~al.}(2015){Woitke}, {Min}, {Pinte}, {Thi}, {Kamp},
  {Rab}, {Anthonioz}, {Antonellini}, {Baldovin-Saavedra}, {Carmona}, {Dominik},
  {Dionatos}, {Greaves}, {G{\"u}del}, {Ilee}, {Liebhart}, {M{\'e}nard},
  {Rigon}, {Waters}, {Aresu}, {Meijerink}, \& {Spaans}}]{2015arXiv151103431W}
{Woitke}, P., {Min}, M., {Pinte}, C., {et~al.} 2015, ArXiv e-prints

\end{thebibliography}

\end{document}